\documentclass[numbook, envcountsect, envcountsame, envcountreset, runningheads, smallextended]{svjour3}

\usepackage{graphicx}        
\usepackage{multicol}        
\usepackage[ruled,vlined]{algorithm2e}
\usepackage{amsmath}
\usepackage{dsfont}
\usepackage{bm}
\usepackage{color}
\usepackage{placeins}
\usepackage[hidelinks]{hyperref}
\usepackage{pbox}
\usepackage{enumerate}
\usepackage{bbm}

\usepackage{ mathrsfs }
\usepackage{multirow}

\usepackage{booktabs}
\usepackage{amsfonts}

\usepackage[table]{xcolor}
\usepackage{booktabs} 

							\newcommand{\R}{\mathds{R}}

							\newcommand{\Prob}{\mathds{P}}
							\newcommand{\Exp}{\mathds{E}}
							
							\newcommand{\N}{\mathds{N}}
							\newcommand{\z}{\mathbf{z}}
							\newcommand{\Z}{\mathbf{Z}}

							\newcommand{\Y}{\mathbf{Y}}
							\newcommand{\y}{\mathbf{y}}
							\newcommand{\X}{\mathbf{X}}

							\newcommand{\x}{\mathbf{x}}

							\newcommand{\bfH}{\mathbf{H}}

							\newcommand{\Ubold}{\mathbf{U}}
							\newcommand{\ubold}{\mathbf{u}}
							
							\newcommand{\bP}{\mathds{P}}
							
							\newcommand{\bR}{\mathds{R}}
							\newcommand{\bE}{\mathds{E}}

							\newcommand{\VaR}{\operatorname{VaR}}
							
							\newcommand{\VaRs}{\mbox{\small VaR}}
							
							\renewcommand*{\i}{{-1}}
							\newcommand{\esssup}{\operatorname{esssup}}
							\newcommand{\essinf}{\operatorname{essinf}}
							\newcommand*{\GPD}{\operatorname{GPD}}
							\newcommand*{\LN}{\operatorname{LN}}
							\newcommand*{\Norm}{\operatorname{N}}
							
							\newcommand*{\Unif}{\operatorname{U}}

							\definecolor{light-gray}{gray}{0.90}
							\definecolor{mid-gray}{gray}{0.50}
							
\usepackage{etoolbox}
\newcommand*{\affaddr}[1]{#1} 
\newcommand*{\affmark}[1][*]{\textsuperscript{#1}}

\definecolor{cadmiumgreen}{rgb}{0.0, 0.42, 0.24}

\usepackage[authormarkup=none]{changes}
\definechangesauthor[name={Rodrigo}, color=orange]{R}
\definechangesauthor[name={Yuri}, color=cadmiumgreen]{Y}
\definechangesauthor[name={Taka}, color=blue]{T}
\definechangesauthor[name={Final}, color=black]{F}
\definechangesauthor[name={Final2}, color=red]{F2}

\begin{document}

\title{Avoiding zero probability events when computing Value at Risk contributions\footnote{This paper has been previously arXiv'ed under the title \emph{Avoiding zero probability events when computing Value at Risk contributions: a Malliavin calculus approach}. The major difference is that we no longer use Malliavin calculus to derive the main result and hence are able to consider a wider class of distributions.}}
\titlerunning{Avoiding zero probability for VaR contributions}

\author{
Takaaki Koike\affmark[1] \and Yuri Saporito\affmark[2] \and
Rodrigo Targino\affmark[2]\footnote[2]{Corresponding author. \email{	rodrigo.targino@fgv.br}}
}

\authorrunning{Koike \and Saporito \and Targino}

\institute{
\affaddr{\affmark[1] Risk Analysis Research Center, The Institute of Statistical Mathematics, Japan. 
}
\\
\affaddr{\affmark[2] School of Applied Mathematics (EMAp), Getulio Vargas Foundation (FGV), Brazil.}
}
\date{\today}
\maketitle

\begin{abstract}
This paper is concerned with the process of risk allocation for a generic multivariate model when the risk measure is chosen as the Value-at-Risk (VaR). We recast the traditional Euler contributions from an expectation conditional on an event of zero probability to a ratio involving conditional expectations whose conditioning events have strictly positive probability.
We derive an analytical form of the proposed representation of VaR contributions for various parametric models.
Our numerical experiments show that the estimator using this novel representation outperforms the standard Monte Carlo estimator in terms of bias and variance.
Moreover, unlike the existing estimators, the proposed estimator is free from hyperparameters under a parametric setting.
\end{abstract}



\section{Introduction}
Let us consider $\X = (X_1,...,X_d)$ the losses (negative of the returns) of $d$ different assets in a portfolio. For a linear portfolio with unitary exposure to each asset, the portfolio-wide loss is defined as $X = \sum_{i=1}^d X_i$. After a risk measure of the portfolio is computed, one is usually interested in understanding how much each asset contributes to the overall portfolio risk, in a process known as risk allocation.

Here we follow the Euler allocation principle, proposed in \cite{tasche1999risk}. This principle arises in different contexts in the literature. For example, in \cite{denault2001coherent} and \cite{kalkbrener2005axiomatic} it is motivated by two (different) sets of axioms, leading to coherent allocation principles (for a relationship between coherent risk measures and coherent capital allocations see \cite{buch2008coherent}). It has also found a wide range of applications, ranging from credit (\cite{tasche1999risk}, \cite{glasserman2005measuring}) and systemic risks (\cite{brownlees2012volatility}, \cite{mainik2014dependence}) to banking and insurance capital \cite{tasche2008capital}.

In this paper we assume the risk measure of interest is the Value-at-Risk (VaR), which is the risk measure used for capital calculation for insurance companies under the Solvency II regulation \cite{solvencyii}. Even though the Basel Committee on Banking Supervision \cite{baselMarketRisk} has changed, in what is known as the Basel 3.5 agreement, the market risk measure from the VaR to the Expected Shortfall (ES), the former is still ubiquitous as a measure of Economic Capital in financial institutions. Even when compared with the ES, the VaR has important practical properties, such as its existence for distributions with infinite mean. For a comprehensive comparison between the risk measures the reader is referred to \cite{embrechts2014academic} and \cite{emmer2015best}.

For the Value-at-Risk with confidence level $\alpha \in (0,1)$, the Euler principle states the \emph{VaR allocation} (also called the \emph{VaR contribution}) to the $i$-th asset is given by
\begin{align} \label{eq:var_allocation}
\mathscr{C}^{\alpha}_i = \Exp[X_i \, | \, X = \VaR_\alpha(X) ],
\end{align}
where $\VaR_\alpha(X)=\operatorname{inf}\{x \in \R: \bP(X\leq x)\geq \alpha\}$. 
Throughout the paper we assume that the losses have a joint density, which implies \eqref{eq:var_allocation}, by the result shown in ~\cite[Section 8.5]{mcneil2015quantitative}; see also, \cite{tasche1999risk}.

The expectation in \eqref{eq:var_allocation} rarely has a closed form solution; with a notable exception being when the losses are jointly Gaussian or, more generally, elliptically distributed \cite{mcneil2015quantitative}. Therefore, approximation strategies need to be used if one is willing to use a generic multivariate loss model.

As the VaR contributions are expectations conditional on an event of probability zero, direct Monte Carlo simulation cannot be used. One strategy, followed by \cite{glasserman2005measuring}, to avoid the zero probability condition in \eqref{eq:var_allocation} is to estimate what we call $\delta$-\emph{allocations}:
\begin{equation}
\mathscr{C}^{\alpha,\delta}_i = \Exp[X_i \, | \, \VaR_{\alpha - \delta}(X)\leq X\leq \VaR_{\alpha + \delta}(X) ], \label{eq:approx_contrib}
\end{equation}
for a sufficiently small $\delta >0$ such that $0 < \alpha-\delta$ and $\alpha+\delta < 1$.
The $\delta$-\emph{estimator} is then defined as a sample mean of i.i.d. samples of $\X\ | \  \{\VaR_{\alpha -\delta}(X)\leq X\leq  \VaR_{\alpha + \delta}(X) \}$.
Despite the simplicity of this estimator, there is a trade-off between $\delta$ and the sample size $N$ in terms of the bias and the standard error; see \cite[Section~2]{koike2019estimation} for details.

To the best of our knowledge, the issue of computing the exact VaR contributions via Monte Carlo simulation has only been addressed in \cite{koike2019estimation} (and its extension in \cite{koike2020markov}) and using Conditional Monte Carlo in \cite{fu2009conditional}. In \cite{koike2019estimation} and \cite{koike2020markov} the authors use a Metropolis--Hastings algorithm to simulate $(X_1,\ldots,X_{d-1})\,|\, \{X = \VaR_\alpha(X))\}$, allied to the fact that $\mathscr{C}_d^{\alpha} = \VaR_\alpha(X) - \sum_{i=1}^{d-1} \mathscr{C}_i^{\alpha}$, in a strategy that resembles the slice sampler employed for the computation of the ES contributions of \cite{targino2015sequential}.
Alternatives to Monte Carlo simulation do exist in the literature. For instance, \cite{gourieroux2000sensitivity}, \cite{tasche2008capital}, \cite{liu2009kernel} compute the VaR contributions by means of a kernel estimator, while \cite{hong2009estimating} derived a batched infinitesimal perturbation analysis (IPA) estimator. More recently, \cite{siller2013measuring} used a technique called Fourier Transform Monte Carlo and \cite{asimit2019efficient} derived a nonparametric procedure which estimate allocations at rate $n^{-1/2}$.

Here we take a different route from existing methods. Instead of directly targeting the distribution of $\X \mid \{X = \VaR_{\alpha}(X)\}$ we build upon the ideas of \cite{FournieMalliavin2} and develop a novel expression for VaR contributions, namely,
\begin{equation}
\mathscr{C}^{\alpha}_i = \frac{B_{iI} + (1 - \alpha) \displaystyle \bE\left[X_i \ \pi_I + \pi_{iI} \ | \ X \geq \VaR_\alpha(X) \right]}{B_I + (1-\alpha)\displaystyle \bE\left[\pi_I \ | \ X \geq \VaR_\alpha(X) \right]},
\label{eq:var_allocation_intro}
\end{equation}
with weights $\pi_I$, $\pi_{iI}$ and boundary terms $B_I$, $B_{iI}$ depending on the distribution of $\X$ and precisely described in Theorems \ref{thm:main} and \ref{thm:extension}.
We call \eqref{eq:var_allocation_intro} the \emph{IBP formula} of VaR contributions where the abbreviation stems from the fact that the formula is derived by adapting the ideas in \cite{FournieMalliavin2} to use Integration By Parts. 
In~\eqref{eq:var_allocation_intro}, the conditioning events of the expectations are enlarged from $\{X =\VaR_\alpha(X)\}$ to $\{X \geq \VaR_\alpha(X)\}$.
The losses $X_i$, $i=1,\dots,d$ are then adjusted by two weights $\pi_{I}$ and $\pi_{i I}$, and boundary terms are added in both numerator and denominator so that the resulting ratio of expectations yields the VaR contributions.
Note that the integrability requirements to use Malliavin calculus would not allow us to consider the wide class of marginal distributions contemplated in our framework. Namely, within Malliavin calculus, the boundary terms would need to vanish.

Since $\bP(X \geq \VaR_\alpha(X))=1-\alpha>0$, a direct Monte Carlo simulation is feasible and one can construct a consistent estimator of $\mathscr{C}^{\alpha}_i$ based on \eqref{eq:var_allocation_intro}.
Through simulation and empirical studies we show that an IBP-based estimator typically outperforms the $\delta$-estimator in terms of the bias and standard error.
Another practical advantage of the IBP-based estimator is that it is free from the choice of hyperparameters in a parametric setting, which, in contrast, is not the case for the existing estimators above. 
For instance, the $\delta$-estimator depends on the choice of $\delta$ (and, as we observe empirically in Section \ref{sec:examples}, the estimate can be very sensitive to it); estimators based on MCMC/Importance Sampling depend on proposal/importance distributions; estimators based on kernels depend on a bandwidth parameter; the batched IPA estimator relies on the choice of a small increment in the derivative approximation; the Fourier Method of \cite{siller2013measuring} depends on the resolution $\Delta$. 
On the other hand, the IBP-based estimator does not require any hyperparameter selection since the direct Monte Carlo method is accessible.

The reader should also note that, besides the fact the expectations in  \eqref{eq:var_allocation_intro}, differently from \eqref{eq:var_allocation}, do not depend on the zero probability event $\{X = \VaR_\alpha(X)\}$, the present conditioning event $\{X \geq \VaR_\alpha(X)\}$ is precisely the same as the one in the ES contributions derived from the Euler principle. 
Therefore, the computationally efficient algorithms derived in \cite{targino2015sequential}, \cite{peters2017bayesian}, \cite{koike2019estimation} and \cite{koike2020markov} could be adapted to estimate the expectations in \eqref{eq:var_allocation_intro}.

	In order to clarify the extent of our contribution, we outline below the workflow from the observed data to the allocation.
\begin{enumerate}[(a)]
	\item Collect loss data $\x_1, \ldots, \x_M \in \R^d$ for $M$ periods and $d$ assets;
	\item Assume a family of parametric models parametrized by $\theta$;
	\item Fit the model to the observations $\x_1, \ldots, \x_M$, leading to an estimated parameter $\widehat{\theta}$;
	\item Check the model fit and, if necessary, return to (b) and reconsider the family of models;
	\item Given the estimated model, estimate the VaR contributions~\eqref{eq:var_allocation}.
\end{enumerate}
	In this paper we assume steps (a) to (d) above have already been dealt with by the analyst, and we address the problem of estimating VaR contributions under the prescribed parametric model. Nevertheless, we stress the fact that the choice of multivariate model for the losses in step (b) may be crucial to the computation of the VaR and, consequently, of its contributions. As a general rule, if not enough data is available (small $N_{obs}$) it may be beneficial to rely on parametric models, even if only for the tails of the distributions. A potential extension of the IBP-based estimator in a nonparametric model is illustrated in Example~\ref{subsec:kde}. As a disclaimer, this work is only concerned with the efficient computation of the VaR contributions given by the Euler method (item (e), above). We also point out that even though the contribution provided by this paper is only applicable to linear portfolios, other sensitivity measures that can be written as conditional expectations may be computed using the same techniques; see Remark~\ref{rem:nonlinear} for details.

The remainder of the paper is organized as follows. In Section \ref{sec:main_results} we state the main result of the paper, i.e. the new expression of the VaR contributions using only expectations conditional on events of positive probability. In Section \ref{sec:examples} we derive a closed form of the IBP formula for several multivariate models.
In Section \ref{sec:numerics} we conduct some simulation studies to assess the efficiency of the estimator based on the proposed IBP formula for the VaR contributions. 
Conclusion and future work are provided in Section \ref{sec:conclusion}.

\section{The IBP formula for VaR contributions} \label{sec:main_results}

In this section we present and discuss the main results of this paper. We start by presenting a generic structure that models need to follow in order for our IBP formula to be applicable. 

\begin{definition}\label{def:model}
A \textit{model} is a set of functions $g_i : D \subset \bR^k \longrightarrow \bR$, with $i \in \{1,\dots,d\}$ and $k \in \N$, and a random vector $\Y$ taking values in the set $D$  such that the losses are given by $X_i = g_i(\Y)$. For a fixed set of indices $I \subset \{1\,\ldots,k\}$, we say that the model is $\mathcal C^s_I$, for $s=1,2,\dots$, if each function $g_i$ has $s$ continuous derivatives almost everywhere (a.e.) with respect to the coordinates in the index $I$. Moreover, we call the random vector $\Y$ \textit{the driving random vector of the model}, or simply the \textit{driver}. Additionally, we define $g(\y) = \sum_{i=1}^d g_i(\y) - \VaR_\alpha(X)$, where $X = \sum_{i=1}^d X_i$.
\end{definition}

Generally, in the actuarial field, a model arises when the loss random vector $\X$ of a portfolio is represented by a map of risk-factors $\Y$.
Another typical situation is when the model $\X$ is simulated by applying $(g_1,\dots,g_d)$ to a rather primitive random vector $\Y$, such as the uniform distribution $\Unif[0,1]^k$.
The uniform driver $\Y\sim \Unif[0,1]^k$ appears in the Knothe--Rosenblatt transform \cite{knothe1957contributions,rosenblatt1952remarks} and more generally vector rank and quantile transforms \cite{fan2020vector}.
Due to its importance, we first state the main result of this paper for the special case of uniform driver. For a $k$-dimensional vector $\bm u$, denote by $\bm u_{-j}$, $j \in \{1,\dots,k\}$, the $(k-1)$-dimensional vector deleting the $j$th component of $\bm u$ from $\bm u$, and by $(\bm u_{-j},v)$, $v \in \bR$, the $k$-dimensional vector whose $j$th component of $\bm u$ is replaced by $v$. For example, if ${\bm u} = (1, 2, 4, 1, 5)$, then ${\bm u}_{-4} = (1, 2, 4, 5)$ and ($\bm u_{-4}, 10) = (1,2,4, 10, 5)$.

\begin{theorem}\label{thm:main}
Consider a set of indices $I\subseteq \{1,\dots,k\}$ and a $\mathcal C^2_I$ model with uniform driver $\Y = \Ubold \sim \Unif[0,1]^k$. Assume that $\partial_j g \neq 0$ a.e. for every $j\in I$. 
Define the following boundary terms
\begin{align*}
B_{iI}&= \sum_{j\in I}\bE\left[\frac{g_i(\Ubold_{-j},u_j)}{\partial_j g(\Ubold_{-j},u_j)}\mathbbm{1}_{\{g(\Ubold_{-j},u_j)\geq 0\}}\Bigg|^{u_j=1}_{u_j=0} \right] \quad\mbox{and}\\
B_I&=\sum_{j\in I}\bE\left[\frac{1}{\partial_j g(\Ubold_{-j},u_j)}\mathbbm{1}_{\{g(\Ubold_{-j},u_j)\geq 0\}}\Bigg|^{u_j=1}_{u_j=0} \right],
\end{align*}
Then, the marginal risk allocation for the VaR risk measure is given by
\begin{align}\label{eq:risk_allocation_pi_uniform}
\mathscr{C}^{\alpha}_i &= \frac{B_{iI} + (1 - \alpha) \displaystyle \bE\left[X_i \ \pi_I + \pi_{iI} \ | \ X \geq \VaR_\alpha(X) \right]}{B_I + (1-\alpha)\displaystyle \bE\left[\pi_I \ | \ X \geq \VaR_\alpha(X) \right]},
\end{align}
with the weights given by 
\begin{align}\label{eq:weights_uniform}
\pi_I = \sum_{j\in I}\frac{\partial_j^2 g(\Ubold)}{(\partial_j g(\Ubold))^2}\quad  \mbox{and}\quad \pi_{iI} = - \sum_{j\in I}\frac{\partial_j g_i(\Ubold)}{\partial_j g(\Ubold)}.
\end{align}
\end{theorem}

Theorem \ref{thm:main} follows directly from Theorem \ref{thm:extension} we present below.
Since formula \eqref{eq:risk_allocation_pi_uniform} is derived by integration by parts (IBP), we call it the \emph{IBP formula} of VaR contributions. 
The formula is beneficial to construct an estimator of VaR contributions since the conditioning event $\{X \geq \VaR_\alpha(X)\}$ in \eqref{eq:risk_allocation_pi_uniform} has positive probability, $1-\alpha$, unlike the original form \eqref{eq:var_allocation}, where 
$\bP(X = \VaR_\alpha(X))=0$ if $X$ is continuous.

\begin{remark}\label{rmk:notation}
For $j \in \{1,\dots,d\}$, we shorthand $B_{ij}$ instead of $B_{iI}$ when $I = \{j\}$. Similarly for $B_j$, $\pi_j$ and $\pi_{ij}$. So, we may write 
$$B_{iI} = \sum_{j\in I} B_{ij}, \quad B_I=\sum_{j\in I}B_{j}, \quad \pi_I = \sum_{j\in I}\pi_{j} \quad\mbox{and}\quad \pi_{iI}=\sum_{j\in I} \pi_{ij}.$$
\end{remark}

\subsection{General driver}

We now consider $X_i=g_i(\Y)$ for some $k$-dimensional general driver $\Y$, which is not necessarily uniformly distributed in $[0,1]^k$. We denote $a_j^\text{Y,U}=\esssup(Y_j)$ and $a_j^\text{Y,L}=\essinf(Y_j)$ as the essential supremum and infimum of $Y_j$, respectively.

\begin{theorem}\label{thm:extension}
Consider a set of indices $I\subseteq \{1,\dots,k\}$ and a $\mathcal C^2_I$ model with general driver $\Y$ admitting a density $f_\Y \in {\mathcal C}^1(\bR^k)$, $k\in \mathbb N$. Assume that $\partial_j g \neq 0$ a.e. for every $j\in I$. 
 Define the following boundary terms
\begin{align*}
B_{iI}&= \sum_{j\in I}\int_{\bR^{k-1}}\frac{g_i(\y_{-j},y_j)f_\Y(\y_{-j},y_j)}{\partial_j g(\y_{-j},y_j)}\mathbbm{1}_{\{g(\y_{-j},y_j)\geq 0\}}\Bigg|^{y_j=a_j^\text{Y,U}}_{y_j=a_j^\text{Y,L}} d \y_{-j}  \quad \mbox{ and }\\
B_{I}&= \sum_{j\in I}\int_{\bR^{k-1}}\frac{f_\Y(\y_{-j},y_j)}{\partial_j g(\y_{-j},y_j)}\mathbbm{1}_{\{g(\y_{-j},y_j)\geq 0\}}\Bigg|^{y_j=a_j^\text{Y,U}}_{y_j=a_j^\text{Y,L}} d \y_{-j}.
\end{align*}
Then, the marginal risk allocation for the VaR risk measure is given by
\begin{align}\label{eq:risk_allocation_pi_general}
\mathscr{C}^{\alpha}_i &= \frac{B_{iI} + (1 - \alpha) \displaystyle \bE\left[X_i \ \pi_I + \pi_{iI} \ | \ X \geq \VaR_\alpha(X) \right]}{B_I + (1-\alpha)\displaystyle \bE\left[\pi_I \ | \ X \geq \VaR_\alpha(X) \right]},
\end{align}
with the weights given by 
\begin{align}\label{eq:weights_general}
\pi_I = \sum_{j\in I}\left(\frac{\partial_j^2 g(\Y)}{(\partial_j g(\Y))^2} 
-\frac{\partial_{j}f_{\Y}(\Y)}{f_{\Y}(\Y)}\frac{1}{\partial_{j} g(\Y)}
\right)\quad \mbox{ and }\quad \pi_{iI} = - \sum_{j\in I}\frac{\partial_j g_i(\Y)}{\partial_j g(\Y)}.
\end{align}
\end{theorem}

\begin{proof}
The proof can be found in Appendix \ref{app:proof_main}. 
\end{proof}

\begin{remark}\label{rem:nonlinear}
As seen in the proof of Theorem \ref{thm:extension}, the particular form of the aggregator $g(\y)$ assumed in Definition \ref{def:model} is not required for the IBP reformulation. 
Therefore, analogous formulas can be derived for more general quantile sensitivities studied in \cite{hong2009estimating}.
Such potential generalizations are left for future research since our particular interest is the VaR contributions of a linear portfolio.
\end{remark}

Through the generalization shown in the theorem above, the weight $\pi_I$ in \eqref{eq:weights_general} is slightly more involved than \eqref{eq:weights_uniform}. 
Examples of drivers and their implication to the weights are provided in the sequel.

\begin{example}[Uniform and normal distributions]\label{ex:weights}
If $\Y\sim \Unif[0,1]^k$, then the same formulas as in Theorem~\ref{thm:main} are found.
If $\Y\sim \Norm(\mathbf{0}_k,I_k)$, then $-\tfrac{\partial_{j}f_{\Y}(\y)}{f_{\Y}(\y)}=y_{j}$ and we conclude 
\begin{align*}
\pi_{j}=\frac{\partial_{j}^2 g(\Y) }{(\partial_{j} g(\Y))^2}+\frac{Y_{j}}{\partial_{j} g(\Y)}.
\end{align*}
\end{example}

\begin{example}[Trivial model] \label{example:trivial:driver}
Let us consider the case with driver $\Y=\X$ (or, equivalently, $g_i(\y) = y_i$). 
Since $\partial_i g_j \equiv 0$, for $j\neq i$, and $\partial_j g \equiv 1$, we find
\begin{align*}
 \pi_{j}&=\frac{\partial_{j}^2 g(\Y) }{(\partial_{j} g(\Y))^2}-\frac{\partial_{j}f_{\Y}(\Y)}{f_{\Y}(\Y)}\frac{1}{\partial_{j} g(\Y)}=-\frac{\partial_{j}f_{\Y}(\Y)}{f_{\Y}(\Y)}\quad\text{and}\\
 \pi_{ij}&=-\frac{\partial_{j} g_i(\Y)}{\partial_{j} g(\Y)}=0.
\end{align*}
This case would be interesting only when $f_\Y$ (which is the same as $f_\X$) is known in closed form and its derivative can be computed explicitly. In general, the boundary terms, $B_{iI}$ and $B_I$, would also have to be computed. Nonetheless, from Theorem \ref{thm:extension}, these terms will vanish whenever the density $f_\Y$ goes to zero at both limits of its support, $a_j^\text{Y,U}$ and $a_j^\text{Y,L}$.
\end{example}

\begin{remark}\label{remark:choice:g}
Despite the attraction of the simple formulas in the trivial model in Example \ref{example:trivial:driver}, there are various reasons for decomposing the target random variable $X_i$ into $g_i$ and $\Y$.
For one, the decomposition enables us to apply the IBP formula \eqref{eq:risk_allocation_pi_general} to wider variety of models.
For example, consider the typical situation in risk management where each loss $X_i$ of the portfolio is represented by a map of risk-factors $\Y$.
When the map involves sums and/or products of $Y_1,\dots,Y_k$, the joint density of $\X$ can be obtained as an integral but it is not analytically available in general.
Formula \eqref{eq:risk_allocation_pi_general}, however, is applicable in this situation as long as the joint density of $\Y$ is known analytically and $g_i$ is sufficiently smooth.
Another reason is to reduce the computational complexity of the formula.
For example, if the formula of the trivial model is applied to an Archimedean copula, then $(d+1)$ derivatives of the generator $\psi$ are required to obtain the weight function $\pi_i$.
On the other hand, only the first and second derivatives of $\psi$ appear in the weight function of the formula based on the decomposition induced by the Marshall-Olkin algorithm; see Section~\ref{sec:ex_archim} and Table~\ref{tab:weights:copulas} for details.
Therefore, one may simplify formula \eqref{eq:risk_allocation_pi_general} by using the knowledge on simulation algorithm of $\X$ from a driver $\Y$ which is typically more tractable than $\X$ itself.
\end{remark}

\begin{example}[Mixture model]\label{example:mixture}
Let $\X = \Y$ where the driver $\Y$ is a mixture of $M$ random vectors $\Y^{(m)}$, each one with density $f_m$, and mixture weights $p_m \in (0,1)$, $m=1,\dots,M$, such that $\sum_{m=1}^M p_m=1$. Therefore, the density of $\X$ can be written as
$$f_{\X}(\x) = \sum_{m=1}^M p_m f_m(\x),$$
which implies, by Example \ref{example:trivial:driver}, that the weights of this model are given by
\begin{align*}
 \pi_{j}= -\frac{\partial}{\partial x_j}\log f_{\X}(\X) = -\frac{\sum_{m=1}^Mp_m  \partial_{j}f_m(\X)}{\sum_{m=1}^M p_m f_m(\X)}\quad\text{and}\quad
 \pi_{ij}=0.
\end{align*}
One case of particular interest, closely related to the kernel density estimator (see Section \ref{subsec:kde}), is when $f_m$ is Gaussian, in which the boundary terms vanish.
\end{example}

When computing the weights in (\ref{eq:weights_general}) one should notice that one of the terms is the negative to what is known as the \textit{score function} in Statistics. More precisely, if we define the \textit{potential energy} associated with the density $f_{\Y}(\y)$ as 
\[E_\Y(\y) = -\log f_{\Y}(\y),\]
the score function of $\Y$ is the gradient of the potential energy
\[\nabla E_\Y(\y) = \left(-\frac{\partial_{1}f_{\Y}(\y)}{f_{\Y}(\y)}, \ldots, -\frac{\partial_{k}f_{\Y}(\y)}{f_{\Y}(\y)} \right).\]	
Note that the weight functions in the trivial model in Example~\ref{example:trivial:driver} are given by this score function.

\begin{remark} \label{rmk:score}
It is straightforward to check that the score function of $\Y$ is invariant under conditioning, that is, for any measurable set $A \subseteq \R^k$, it holds that
$$
\nabla E_\Y(\y) = \nabla E_{\Y \mid \{\Y \in A\}}(\y)\quad \text{for all } \y \in A\backslash \partial A,
$$
where $\partial A$ is the boundary of $A$.
\end{remark}

\section{Weights and boundary terms for parametric copula families} \label{sec:examples}

\begin{table}[t!]
	\begin{center}
		\begin{tabular}{ll}
			\toprule
			\textbf{Copula family}           & \textbf{Weights and boundary terms} \\
			\midrule
			\rowcolor[gray]{0.95}
			Independence Copula & 
			$
			B_{iI}=\sum_{j \in I} b_j^{\text{U}} \bE[X_i \mathbbm{1}_{\{a_j^{\text{U}} + X_{-j} \geq \VaRs_\alpha(X)\}}] -  \sum_{j \in I}b_j^{\text{L}}\bE[ X_i \mathbbm{1}_{\{a_j^{\text{L}} + X_{-j}\geq \VaRs_\alpha(X)\}}]$,\\ [10pt] \rowcolor[gray]{0.95}
			& 
			$B_{I}= \sum_{j \in I}b_j^{\text{U}} \bP(a_j^{\text{U}} + X_{-j}\geq \VaR_\alpha(X)) -  \sum_{j \in I}b_j^{\text{L}} \bP(a_j^{\text{L}} + X_{-j}\geq \VaR_\alpha(X))$,\\[10pt] \rowcolor[gray]{0.95}
			&  
			$ \pi_{iI} =0 \quad\text{and}\quad\pi_I= \sum_{j \in I} E_j'(X_j).
			$\\
			\midrule
			Archimedean Copula &
			$B_{iI}=-\sum_{j \in I}\frac{b_{j}^\text{U}}{\psi'(0)}\bE[X_i V
			\mathbbm{1}_{\{a_{j}^\text{U}+X_{-j}\geq \VaR_\alpha(X)\}}
			]+\sum_{j \in I}b_{j}^\text{L}\bE[X_i V r_\psi(V)
			\mathbbm{1}_{\{a_{j}^\text{L}+X_{-j}\geq \VaR_\alpha(X)\}}
			],$\\[10pt]
			&
			$B_{I}=-\sum_{j \in I}\frac{b_{j}^\text{U}}{\psi'(0)}\bE[V
			\mathbbm{1}_{\{a_{j}^\text{U}+X_{-j}\geq \VaR_\alpha(X)\}}
			]+\sum_{j \in I}b_{j}^\text{L}\bE[V r_\psi(V)
			\mathbbm{1}_{\{a_{j}^\text{L}+X_{-j}\geq \VaR_\alpha(X)\}}
			]$,\\[10pt]
			&
			$\pi_{iI}\equiv 0,\quad\text{and}\quad
			\pi_{I}=\sum_{j \in I}
			E'_{j}(X_j)+\gamma_{j}(\Ubold;\psi)f_{j}(X_j)$,\\[10pt]
			&
			\text{where}\quad $\gamma_{j}(\ubold;\psi)=\frac{\psi''(\eta_{j}(\ubold))}{(\psi'(\eta_{j}(\ubold)))^2}+
			\frac{Q_V(u_k)}{\psi'(\eta_{j}(\ubold))}$.
			\\
			\midrule
			\rowcolor[gray]{0.95}
			Survival Archimedean &
			$B_{iI}=\sum_{j\in I}\frac{b_{j}^\text{L}}{\psi'(0)}\bE[X_i V
			\mathbbm{1}_{\{a_{j}^\text{L}+X_{-j}\geq \VaR_\alpha(X)\}}
			]-\sum_{j\in I}b_{j}^\text{U}\bE[X_i V r_\psi(V)
			\mathbbm{1}_{\{a_{j}^\text{U}+X_{-j}\geq \VaR_\alpha(X)\}}
			]$,\\[10pt] \rowcolor[gray]{0.95}
			\smash{\raisebox{15pt}{Copula}}  & 
			$B_{I}=\sum_{j\in I}\frac{b_{j}^\text{L}}{\psi'(0)}\bE[V
			\mathbbm{1}_{\{a_{j}^\text{L}+X_{-j}\geq \VaR_\alpha(X)\}}
			]-\sum_{j\in I}b_{j}^\text{U}\bE[V r_\psi(V)
			\mathbbm{1}_{\{a_{j}^\text{U}+X_{-j}\geq \VaR_\alpha(X)\}}
			]$,\\[10pt] \rowcolor[gray]{0.95}
			& 
			$\pi_{iI}\equiv 0\quad\text{and}\quad
			\pi_{I}=\sum_{j\in I}
			E'_{j}(X_j)-\gamma_{j}(\Ubold;\psi)f_{j}(X_j)$,
			\\[10pt] \rowcolor[gray]{0.95}
			& 
			\text{where}\quad $\gamma_{j}(\ubold;\psi)=\frac{\psi''(\eta_{j}(\ubold))}{(\psi'(\eta_{j}(\ubold)))^2}+
			\frac{Q_V(u_k)}{\psi'(\eta_{j}(\ubold))}$.
			\\
			\midrule
			Elliptical Copula & 
			$B_{iI}=B_{I}=0$,\\[10pt]&
			$\pi_{iI}\equiv 0\quad\text{and}\quad
			\pi_I = \sum_{j\in I} E'_{j}(X_j)+\gamma_{j}(\Y;\xi_d,P)f_{j}(X_j)$,			
			\\[10pt] & 
			\text{where}\quad $\gamma_{j}(\y;\xi_d,P)=\frac{f'_{Y_{j}}(y_{j})}{f_{Y_{j}}^2(y_{j})}-\frac{\xi'_d\left(\frac{1}{2}\y^{\top}P^\i\y\right)}{\xi_d\left(\frac{1}{2}\y^{\top}P^\i\y\right)}\frac{\y^{\top}\mathbf{q}_{j}}{f_{Y_{j}}(y_{j})}.$
			\\
			\bottomrule
		\end{tabular}
	\end{center}
	\caption{
		Weights and boundary terms for parametric copula families with marginal distributions $X_i$, $i=1,\dots,d$.
		The $i$th margin has a density $f_i$, potential energy $E_i(x)=-\log f_i(x)$, essential supremum $a_j^\text{U}=\esssup(X_i)$ with $b_j^\text{U}=f_i(a_i^\text{U})$ and essential infimum $a_j^\text{L}=\essinf(X_i)$ with $b_j^\text{L}=f_i(a_i^\text{L})$.
		For a (survival) Archimedean copula, $\psi$ is a generator function, $V\sim F_\psi= \mathcal{LS}^{-1}(\psi)$ with $Q_V=F_\psi^{-1}$ and $\eta_j(\ubold)=-\log(u_j)/Q_V(u_{d+1})$, $\ubold \in \bR^{d+1}$.
		For an elliptical copula, $\xi_d$ is a density generator, $P$ is a correlation matrix and $Q=(\mathbf{q}_1,\dots,\mathbf{q}_d)=P^\i$.
	} \label{tab:weights:copulas}
\end{table}

In this section we consider different choices of $\Y$ and $g_i$, which imply different distributions for $\X$. We then derive the weights and boundary terms for each given model. 

For $i=1,\dots,d$, let $X_i \sim F_i$ be the marginal distribution and let $a_i^\text{U}=\esssup(X_i)$ and $a_i^\text{L}=\essinf(X_i)$ be the essential supremum and infimum of $X_i$, respectively.
Throughout this section, we assume the following conditions on $F_1,\ldots,F_d$ :
\begin{enumerate}[leftmargin=*]
\item[(MC1)] $F_i$ is continuous and strictly increasing, and admits a density $f_i \in {\mathcal C}^1$,\\
\item[(MC2)] $b_i^{\text{U}}=f_i(a_i^{\text{U}}):=\lim_{u\uparrow 1}f_i(F_i^{\i}(u))<\infty$ and $b_i^{\text{L}}=f_i(a_i^{\text{L}}):=\lim_{u\downarrow 0}f_i(F_i^{\i}(u))<\infty$.
\end{enumerate}
Notice that, by the ``Marginal Conditions" (MC1) and (MC2), we must have $0 < f_i(X_i) < \infty$ almost surely.
Moreover, we denote the potential energy associated with the density $f_j$ by $E_j(x)=-\log f_j(x)$.
The copula of $\X$ is denoted by $C$, and the copula density is given by $c=\partial_{(1,\dots,d)}C$ provided it exists. We then analyze the cases for Independence, Archimedean and Elliptical copulas.
Several examples of models (i.e. choices of $g_i$), their correspondent weights and boundary terms are studied in the following sections, and summarized in Tables \ref{tab:weights:copulas}, \ref{tab:weights:copulas_2} and \ref{tab:weights:margins}.

\begin{table}[t!]
	\begin{center}
		\begin{tabular}{ll}
			\toprule
			\textbf{Copula}           & \textbf{Parameters} \\
			\midrule
			Clayton Copula & 
			$\gamma_j(\ubold;\psi)=\frac{-\vartheta(Q_V(u_k)-\log(u_j))+\vartheta+1}{\psi(\eta_j(\ubold))},$\\[10pt]
			&
			where $\psi(t) = (1+t)^{-1 / \vartheta}$, $\psi'(0)=-1/\vartheta$ and  $r_{\psi}(v)=0.$\\
			\midrule
			\rowcolor[gray]{0.95}
			Gumbel Copula & 
			$\gamma_j(\ubold;\psi)=\frac{-\vartheta Q_V(u_k) \eta_j(\ubold)^{1-1/\vartheta}+(\vartheta-1)\eta_j(\ubold)^{-1/\vartheta}+1
			}{\psi(\eta_j(\ubold))},$ \\[10pt] \rowcolor[gray]{0.95}
			& 
			where $\psi(t) = e^{-t^{1/\vartheta}}$, $\psi'(0)=-\infty$ and  $r_{\psi}(v)=0.$\\
			\midrule
			Gaussian Copula & 
			$\gamma_i(\y;\xi_d,P)=\sqrt{2\pi}(\y^{\top}\mathbf{q}_j-y_j)\exp\left(\frac{1}{2}y_j^2\right)$,\\[10pt]
& where $\xi_d(t)=\exp(-t)$.\\
			\midrule
			\rowcolor[gray]{0.95}
			$t$ Copula & 
			$\gamma_i(\y;\xi_d,P)=-\frac{\nu+1}{\nu}\frac{y_j}{r}\Big(1+\frac{y_j^2}{\nu}\Big)^{\frac{\nu-1}{2}}+\frac{d+\nu}{\nu}
\frac{\y^{\top}\mathbf{q}_{j}}{r}
\left(1+\frac{\y^{\top}P^\i \y}{\nu}\right)^{-1}\Big(1+\frac{y_j^2}{\nu}\Big)^{\frac{\nu+1}{2}}$,\\[10pt]
\rowcolor[gray]{0.95} & where
 $\xi_d(t)=\left(1+\frac{2t}{\nu}\right)^{-\frac{d+\nu}{2}}$ and $r=\frac{\Gamma\left(\frac{\nu+1}{2}\right)}{\sqrt{\nu\pi}\Gamma\left(\frac{\pi}{2}\right)}$.\\
			\bottomrule
		\end{tabular}
	\end{center}
	\caption{Typical parametric copula families to be used in Table \ref{tab:weights:copulas}.
	} \label{tab:weights:copulas_2}
\end{table}

\subsection{Independence copula} \label{sec:ex_indep}

Consider the case $C(\ubold) = \Pi(\ubold)=\operatorname{\Pi}_{j=1}^d u_j$.
By taking $k=d$, we consider the model
\begin{align*}
g_i(\ubold)=F_i^{{\i}}(u_i)\quad\text{and}\quad g(\ubold)=\sum_{m=1}^d F_m^{{\i}}(u_m)-\VaR_\alpha(X).
\end{align*}
Then, for $X_i=g_i(\Ubold)$, $i=1,\dots,d$, $\Ubold \sim \Unif[0,1]^k$, the random vector $\X=(X_1,\dots,X_d)$ has margins $F_1,\dots,F_d$ and a copula $\Pi$.
We now derive the boundary terms and weights of this model as in Theorem~\ref{thm:main}.

For $j \in \{1,\dots,d\}$, we have that 
\begin{align*}
\partial_j  g(\ubold)&=\partial_j F_j^{\i}(u_j)=\frac{1}{f_j(F_j^{\i}(u_j))},\\
\partial_j^2  g(\ubold)&=-\frac{f_j'(F_j^{\i}(u_j))}{f_j(F_j^{\i}(u_j))^3}.
\end{align*} 
Then, for $j\neq i$, we have
\begin{align*}
\pi_j &= \frac{\partial_j^2  g(\Ubold)}{(\partial_j  g(\Ubold))^2} = -\frac{f_j'(F_j^{\i}(U_j))}{f_j(F_j^{\i}(U_j))} = E_j'(X_j),\\
\pi_{ij} &= -\frac{\partial_j  g_i(\Ubold)}{\partial_j  g(\Ubold)} = 0,
\end{align*}
where $E_j = -\log f_j$. Let us now analyze the boundary terms. Defining $X_{-j}=\sum_{m\neq j}X_m$, we have 
\begin{align*}
B_{ij}&=\int_{[0,1]^{k-1}} f_j(F_j^{\i}(u_j))F_i^{{\i}}(u_i) \mathbbm{1}_{\{g(\ubold))\geq 0\}}\Bigg|^{u_j=1}_{u_j=0}d \ubold_{-j} \\
&= b_j^{\text{U}} \int_{[0,1]^{k-1}} F_i^{{\i}}(u_i) \mathbbm{1}_{\{a_j^{\text{U}} + g_{-j}(\ubold_{-j})\geq 0\}}d \ubold_{-j} - b_j^{\text{L}}\int_{[0,1]^{k-1}}  F_i^{{\i}}(u_i) \mathbbm{1}_{\{a_j^{\text{L}} + g_{-j}(\ubold_{-j})\geq 0\}}d \ubold_{-j},\\
&= b_j^{\text{U}} \bE[X_i \mathbbm{1}_{\{a_j^{\text{U}} + X_{-j} \geq \VaRs_\alpha(X)\}}] - b_j^{\text{L}}\bE[ X_i \mathbbm{1}_{\{a_j^{\text{L}} + X_{-j}\geq \VaRs_\alpha(X)\}}],\\
B_j&=\int_{[0,1]^{k-1}} f_j(F_j^{\i}(u_j)) \mathbbm{1}_{\{g(\ubold))\geq 0\}}\Bigg|^{u_j=1}_{u_j=0}d \ubold_{-j}\\
&=  b_j^{\text{U}} \int_{[0,1]^{k-1}} \mathbbm{1}_{\{a_j^{\text{U}} + g_{-j}(\ubold_{-j})\geq 0\}}d \ubold_{-j} -  b_j^{\text{L}} \int_{[0,1]^{k-1}} \mathbbm{1}_{\{a_j^{\text{L}} + g_{-j}(\ubold_{-j})\geq 0\}}d \ubold_{-j},\\
&=  b_j^{\text{U}} \bP(a_j^{\text{U}} + X_{-j}\geq \VaR_\alpha(X)) -  b_j^{\text{L}} \bP(a_j^{\text{L}} + X_{-j}\geq \VaR_\alpha(X)).
\end{align*}
Therefore, the weights and the boundary terms in Theorem~\ref{thm:main} are given by
\begin{align*}
B_{iI}&=\sum_{j \in I}B_{ij}=\sum_{j \in I} b_j^{\text{U}} \bE[X_i \mathbbm{1}_{\{a_j^{\text{U}} + X_{-j} \geq \VaRs_\alpha(X)\}}] -  \sum_{j \in I}b_j^{\text{L}}\bE[ X_i \mathbbm{1}_{\{a_j^{\text{L}} + X_{-j}\geq \VaRs_\alpha(X)\}}],\\
B_{I}&=\sum_{j \in I}B_j= \sum_{j \in I}b_j^{\text{U}} \bP(a_j^{\text{U}} + X_{-j}\geq \VaR_\alpha(X)) -  \sum_{j \in I}b_j^{\text{L}} \bP(a_j^{\text{L}} + X_{-j}\geq \VaR_\alpha(X)),\\
\pi_{iI}&=\sum_{j \in I}\pi_{ij} =0\quad\text{and}\quad
\pi_I=\sum_{j \in I}\pi_j = \sum_{j \in I} E_j'(X_j).
\end{align*}

The boundary terms typically appear when the marginal distributions are one-sided (e.g., exponential and generalized Pareto), and vanish when $b_j^\text{U}=b_j^\text{L}=0$; see Table \ref{tab:weights:margins} for particular distributions.

\begin{table}[t!]
\begin{center}
\begin{tabular}{lllr}
\toprule
\textbf{Distribution}   & \textbf{Parameters}        & \textbf{Potential Energy $E_j'(X_j)$}           & \textbf{Boundaries}                                                 \\
\midrule
\rowcolor[gray]{0.95}

Gaussian &  $\Norm(0,\sigma_j^2)$ & $\frac{X_j}{\sigma_j^2}$ &
 $b_j^{\text{U}} = b_j^{\text{L}} = 0$, $a_j^{\text{U}} =+\infty$, $a_j^{\text{L}} = -\infty$\\[10pt]
\midrule
Skew $t$ \cite{fernandez1998bayesian} & $Skt(\nu_j,\gamma_j)$ & $\gamma_j^{2s_j}\frac{(\nu_j+1)X_j}{\nu_j + \gamma_j^{2s_j} X_j^2}$, where $s_j = -\mbox{sign}(X_j)$ &
 $b_j^{\text{U}} = b_j^{\text{L}} = 0$, $a_j^{\text{U}} =+\infty$, $a_j^{\text{L}} = -\infty$\\[10pt]
\rowcolor[gray]{0.95}
\midrule
Generalized Pareto & $\GPD(\xi_j, \beta_j)$ & $\frac{1+\xi_j}{\beta_j}\left(1 + \xi_j \frac{X_j}{ \beta_j}\right)^{-1}$ &
$b_j^{\text{U}} = 0$, $b_j^{\text{L}} = 1/\beta_j$, $a_j^{\text{U}} =+\infty$, $a_j^{\text{L}} = 0$\\[10pt]
\midrule
Exponential  & $\operatorname{Exp}(\lambda_j)$ & $ \lambda_j$ &
$b_j^{\text{U}} = 0$, $b_j^{\text{L}} = \lambda_j$, $a_j^{\text{U}} =+\infty$, $a_j^{\text{L}} = 0$\\[10pt]
\midrule

\rowcolor[gray]{0.95}
Log-normal & $\LN(\mu_j, \sigma_j)$ & $ \frac{1}{X_j}\left(\frac{\log X_j - \mu_j}{\sigma_j^2} + 1\right)$ & 
 $b_j^{\text{U}} = b_j^{\text{L}} = 0$, $a_j^{\text{U}} =+\infty$, $a_j^{\text{L}} = 0$\\[10pt]
\midrule
Gamma & $\Gamma(\alpha_j,\beta_j)$, $\alpha_j>1$ & $\left(\beta_j - \frac{\alpha_j - 1}{X_j} \right) $ &   $b_j^{\text{U}} = b_j^{\text{L}} = 0$, $a_j^{\text{U}} =+\infty$, $a_j^{\text{L}} = 0$\\

\bottomrule
\end{tabular}
\end{center}
\caption{Derivative of the potential energy $E_j(x)=-\log f_j(x)$ for different marginal distribution $X_j$ having a density $f_j$ along with the essential supremum $a_j^\text{U}=\esssup(X_i)$ with $b_j^\text{U}=f_j(a_j^\text{U})$ and essential infimum $a_j^\text{L}=\essinf(X_j)$ with $b_j^\text{L}=f_j(a_j^\text{L})$.
} \label{tab:weights:margins}
\end{table}

\subsection{Archimedean copula}\label{sec:ex_archim}

Let us now consider a random vector $\X = (X_1,\ldots,X_d)$ with the joint distribution given by 
\begin{equation}
F(x_1,\ldots,x_d) = C_{\psi}(F_1(x_1), \ldots, F_d(x_d)),
\label{eq:joint_dist}
\end{equation}
where $C_\psi(\ubold)=\psi\left(\sum_{j=1}^d\psi^{-1}(u_j)\right)$ is an Archimedean copula with completely monotone generator $\psi\in {\mathcal C}^\infty(\bR_{+})$ taking values in $[0,1]$, that is, $\psi$ is continuous, decreasing, strictly decreasing on $[0,\inf\{t:\psi(t)=0\}]$, $\psi(0)=1$, $\psi(\infty):=\lim_{t\rightarrow \infty}\psi(t)=0$, and $(-1)^k\frac{d^k}{d t^k}\psi(t)\geq 0$ for all $k \in \mathbb N$.
We denote by $F_\psi = \mathcal{LS}^{-1}(\psi)$ the inverse Laplace-Stieltjes transform of $\psi$.

Following the Marshall-Olkin algorithm \cite{marshall1988families}, a sample from $(X_1,\dots,X_d)$ following \eqref{eq:joint_dist} can be generated as follows:
\begin{enumerate}[(a)]
\item sample $V \sim F_\psi = \mathcal{LS}^{-1}(\psi)$;
\item sample $U_i \stackrel{iid}{\sim} \Unif[0,1]$, $i=1,\ldots,d$; and
\item define $X_i = F_i^\i\left(\psi( - \log(U_i)/V )\right)$, $i=1,\ldots,d$.
\end{enumerate}

In order to write each $X_i$ as a function of i.i.d. uniform random variables, we first note that the Marshall-Olkin algorithm needs $k = d+1$ i.i.d. uniform random variables $U_1,\ldots, U_{k-1}$ from item (b) and $U_k$, which is such that $V = Q_V(U_k)$ with $Q_V = F_\psi^{-1}$ being the quantile of $F_\psi$. Therefore, we write the random variable $X_i$ as
\[ X_i = g_i(\Ubold)=F_i^{-1}\left(\psi( \eta_i(\Ubold))\right), \]
where we write
$$\eta_i(\ubold)=-\frac{\log(u_i)}{Q_V(u_k)},\quad \ubold \in (0,1)^k
$$
for notational convenience.
In addition to (MC1) and (MC2), assume that $\psi$ satisfies the following ``Archimedean Conditions":
\begin{enumerate}[leftmargin=*]
\item[(AC1)] $\psi$ is strictly decreasing on $\bR_{+}$, and
\item[(AC2)] $\bE[V^2]<\infty$ and $\bE[V^2r_\psi^2(V)]<\infty$, where $r_\psi(v)=\displaystyle\lim_{t\downarrow 0}\frac{t}{\psi'\left(-\log(t)/v\right)}$, $v >0$.
\end{enumerate}
We now derive the boundary terms and weights in Theorem~\ref{thm:main} for this model. Since
\begin{align*}
\partial_{j}\psi(\eta_i(\ubold))=\begin{cases}
\displaystyle-\psi'(\eta_i(\ubold))\frac{1}{u_i Q_V(u_k)},&\text{ if } j=i,\\ \\
\displaystyle\psi'(\eta_i(\ubold))\log(u_i)\frac{Q'_V(u_k)}{Q_V^2(u_k)},&\text{ if } j=k,\\ \\
0, & \text{ if } j\neq i,k,
\end{cases}
\end{align*}
we have, for $j \neq k$, that
\begin{align*}
\partial_{j}g(\ubold)&=\sum_{m=1}^d \partial_{j}g_m(\ubold)=
-\frac{\psi'(\eta_{j}(\ubold))}{f_{j}(g_{j}(\ubold))u_{j}Q_V(u_k)},
\end{align*}
which is non-zero a.e.\ by (AC1).
Next, we have, for $H_{ij}(\ubold) = g_i(\ubold)/\partial_j g(\ubold)$, that 
\begin{align*}
H_{ij}(\ubold)=-\frac{g_i(\ubold)f_{j}(g_{j}(\ubold))u_{j}Q_V(u_k)}{\psi'(\eta_{j}(\ubold))}.
\end{align*}
To study the boundary terms, we first notice that
$\lim_{u_{j}\uparrow 1}\eta_j(\ubold)=0$ and $\lim_{u_{j}\downarrow 0}\eta_j(\ubold)=\infty$, and thus
\begin{align*}
\lim_{u_{j}\uparrow 1}g_j(\ubold)&=\lim_{u\uparrow 1}F_j^\i(u)=\esssup(X_j)=a_j^\text{U},\\
\lim_{u_{j}\downarrow 0}g_j(\ubold)&=\lim_{u\downarrow 0}F_j^\i(u)=\essinf(X_j)=a_j^\text{L}.
\end{align*}
Therefore, we have that
\begin{align*}
\lim_{u_{j}\uparrow 1}H_{ij}(\ubold)
=-\frac{b_{j}^\text{U}g_i(\ubold_{-j},1)Q_V(u_k)}{\psi'(0)}
\end{align*}
and
\begin{align*}
\lim_{u_{j}\downarrow 0}H_{ij}(\ubold)&=-b_{j}^\text{L}g_i(\ubold_{-j},0)Q_V(u_k)\lim_{u_{j}\downarrow 0}\frac{u_{j}}{\psi'(\eta_{j}(\ubold))}
=-b_{j}^\text{L}g_i(\ubold_{-j},0)Q_V(u_k)
r_\psi(Q_V(u_k)).
\end{align*}
Note that $\psi'(0)<0$ since $\psi$ is completely monotone.

We then have that
\begin{align*}
B_{ij}=-\frac{b_{j}^\text{U}}{\psi'(0)}\bE[X_i V
\mathbbm{1}_{\{a_{j}^\text{U}+X_{-j}\geq \VaR_\alpha(X)\}}
]+b_{j}^\text{L}\bE[X_i V r_\psi(V)
\mathbbm{1}_{\{a_{j}^\text{L}+X_{-j}\geq \VaR_\alpha(X)\}}
],
\end{align*}
which is finite by (AC2) and Cauchy--Schwarz inequality.
To compute the weights, we have that $\tilde \pi_{ij}\equiv 0$ by its definition.
By calculation, we find
\begin{align*}
\partial_{j}^2g(\ubold)&=
-\frac{\psi'(\eta_{j}(\ubold))}{u_{j}Q_V(u_k)}\left(\frac{-1}{f_{j}^2(g_{j}(\ubold))}\right)
\partial_{j}f_{j}(g_{j}(\ubold))
\\
&\hspace{10mm}-\frac{1}{f_{j}(g_{j}(\ubold))u_{j}Q_V(u_k)}\psi''(\eta_{j}(\ubold))\partial_{j}\eta_{j}(\ubold)\\
&\hspace{10mm}-\frac{\psi'(\eta_{j}(\ubold))}{f_{j}(g_{j}(\ubold))Q_V(u_k)}\left(\frac{-1}{u_{j}^2}\right)\\
&=
\frac{\psi'(\eta_{j}(\ubold))}{f_{j}^2(g_{j}(\ubold))u_{j}Q_V(u_k)}
f'_{j}(g_{j}(\ubold))\frac{1}{f_{j}(g_{j}(\ubold))}\psi'(\eta_{j}(\ubold))\left(\frac{-1}{u_{j}Q_V(u_k)}\right)\\
&\hspace{10mm}-\frac{\psi''(\eta_{j}(\ubold))}{f_{j}(g_{j}(\ubold))u_{j}Q_V(u_k)}\left(\frac{-1}{u_{j}Q_V(u_k)}\right)\\
&\hspace{10mm}+\frac{\psi'(\eta_{j}(\ubold))}{f_{j}(g_{j}(\ubold))u_{j}^2Q_V(u_k)}.
\end{align*}
Since
\begin{align*}
(\partial_{j}g(\ubold))^2=\frac{(\psi'(\eta_{j}(\ubold)))^2}{f_{j}^2(g_{j}(\ubold))u_{j}^2Q_V^2(u_k)},
\end{align*}
we have that
\begin{align*}
\pi_{j}(\ubold)&=-\frac{{f'}_{j}(g_{j}(\ubold))}{f_{j}(g_{j}(\ubold))}
+\frac{\psi''(\eta_{j}(\ubold))}{(\psi'(\eta_{j}(\ubold)))^2}f_{j}(g_{j}(\ubold))+
\frac{Q_V(u_k)}{\psi'(\eta_{j}(\ubold))}f_{j}(g_{j}(\ubold))\\
&=
E'_{j}(g_{j}(\ubold))+\gamma_{j}(\ubold;\psi)f_{j}(g_{j}(\ubold)),
\end{align*}
where
\begin{align*}
\gamma_{j}(\ubold;\psi)=\frac{\psi''(\eta_{j}(\ubold))}{(\psi'(\eta_{j}(\ubold)))^2}+
\frac{Q_V(u_k)}{\psi'(\eta_{j}(\ubold))}.
\end{align*}
Consequently, we derive the boundary terms and the weights given by
\begin{align*}
B_{iI}&=-\sum_{j \in I}\frac{b_{j}^\text{U}}{\psi'(0)}\bE[X_i V
\mathbbm{1}_{\{a_{j}^\text{U}+X_{-j}\geq \VaR_\alpha(X)\}}
]+\sum_{j \in I}b_{j}^\text{L}\bE[X_i V r_\psi(V)
\mathbbm{1}_{\{a_{j}^\text{L}+X_{-j}\geq \VaR_\alpha(X)\}}
],\\
B_{I}&=-\sum_{j \in I}\frac{b_{j}^\text{U}}{\psi'(0)}\bE[V
\mathbbm{1}_{\{a_{j}^\text{U}+X_{-j}\geq \VaR_\alpha(X)\}}
]+\sum_{j \in I}b_{j}^\text{L}\bE[V r_\psi(V)
\mathbbm{1}_{\{a_{j}^\text{L}+X_{-j}\geq \VaR_\alpha(X)\}}
],\\
\pi_{iI}&\equiv 0\quad\text{and}\quad
\pi_{I}=\sum_{j \in I}
E'_{j}(X_j)+\gamma_{j}(\Ubold;\psi)f_{j}(X_j).
\end{align*}
Note that the first term of $\pi_{I}$ is the weight for the case when $X_1,\dots,X_d$ are independent, and the function $\gamma_j$ is independent of the marginal distributions.
We next focus on particular cases of Archimedean copulas and their survival counterparts.

\subsubsection{Clayton copula} \label{sec:ex_clayton}

For $\psi(t) = (1+t)^{-1 / \vartheta}$ with $\vartheta \in (0,\infty)$, one gets the Clayton family of copulas, where $V \sim \Gamma(1/\vartheta,1)$. 
Then $\psi'_\vartheta(t)=-\frac{1}{\vartheta}(1+t)^{-1/\vartheta-1}< 0$, and thus Condition (AC1) holds.
Moreover, we have that $\psi'_\vartheta(0)=-1/\vartheta< 0$ and that
\begin{align*}
r_{\psi_\vartheta}(v)&=\lim_{t\downarrow 0}\frac{t}{-\frac{1}{\vartheta}\left(1-\frac{\log(t)}{v}\right)^{-\frac{1}{\vartheta}-1}}=-\vartheta\lim_{s\rightarrow \infty}\frac{\exp(-vs)}{(1+s)^{-1/\vartheta-1}}=0.
\end{align*}
Therefore, we have $B_{ij}\equiv 0$ if $b_j^\text{U}=0$.
Since $V\sim \Gamma\left(1/\vartheta,1\right)$, condition (AC2) holds.
Finally, we have that
\begin{align*}
\gamma_j(\ubold;\psi)=\frac{-\vartheta(Q_V(u_k)-\log(u_j))+\vartheta+1}{\psi_\vartheta(\eta_j(\ubold))}.
\end{align*}

\subsubsection{Gumbel copula}

In this case, $\psi(t) = e^{-t^{1/\vartheta}}$ for $\vartheta \in [1,+\infty)$, where $V \sim S(1/\vartheta,1,\left(\cos(\tfrac{\pi}{2\vartheta})\right)^\vartheta,0;1)$, Nolan's 1-parametrization of stable distribution \cite{nolan:2018}.
Condition (AC2) holds for $V$ since $\vartheta > 1/2$.
Moreover, we have $$\psi'_\vartheta(t)=-\frac{1}{\vartheta}t^{-\left(1-1/\vartheta\right)}\exp(-t^{1/\vartheta})<0,$$ and thus Condition (AC1) holds.
Moreover, we have that $\psi'_\vartheta(0)=-\infty$ and that
\begin{align*}
r_{\psi_\vartheta}(v)&=\lim_{s\rightarrow \infty}\frac{\exp(-vs)}{\psi'_\vartheta(v)}=
-\vartheta\lim_{s\rightarrow \infty}\frac{\exp(-vs)}{s^{-\left(1-1/\vartheta\right)}\exp(-s^{1/\vartheta})}\\
&=
-\frac{\vartheta}{v}\lim_{s\rightarrow \infty}\frac{vs}{\exp(vs)}\frac{\exp(s^{1/\vartheta})}{s^{1/\vartheta}}=0.
\end{align*}
Therefore, we have $B_{ij}\equiv 0$ if $b_j^\text{L}=0$.
Finally, we find
\begin{align*}
\gamma_j(\ubold;\psi)=\frac{-\vartheta Q_V(U_k) \eta_j(\ubold)^{1-1/\vartheta}+(\vartheta-1)\eta_j(\ubold)^{-1/\vartheta}+1
}{\psi_\vartheta(\eta_j(\ubold))}.
\end{align*}

\subsubsection{Survival Archimedean copulas}
When $C:[0,1]^d \rightarrow [0,1]$ is a copula and $\mathbf{\mathcal{U}}=(\mathcal{U}_1,\dots,\mathcal{U}_d) \sim C$, its survival copula is defined as $\widetilde{C}(\ubold) = \overline{C}(1-u_1,\ldots,1-u_d)$, where $\overline{C}(\ubold) = \Prob(\mathcal{U}_1>u_1,\ldots,\mathcal{U}_d>u_d)$. In other words, the survival copula is the copula associated with the survival function of $\mathbf{\mathcal{U}}$. Therefore, to sample $\widetilde{\mathbf{\mathcal{U}}} \sim \widetilde{C}$ one can simply sample $\mathbf{\mathcal{U}} \sim C$ and define $\widetilde{\mathbf{\mathcal{U}}}_i = 1-\mathbf{\mathcal{U}}_i$, for $i=1,\ldots, d$.
Following the Marshall-Olkin algorithm, we set $k=d+1$ and write
\begin{align*}
X_i = \tilde g_i(\ubold)\quad\text{where}\quad \tilde g_i(\ubold)=F_i^\i\left(1-\psi\left(-\frac{\log(u_i)}{Q_V(u_k)}\right)\right)=F_i^\i\left(1-\psi\left(\eta_i(\ubold)\right)\right).
\end{align*}
A simple calculation shows that the weights and the boundary terms of the corresponding survival copula are given by
\begin{align*}
B_{iI}&=\sum_{j\in I}\frac{b_{j}^\text{L}}{\psi'(0)}\bE[X_i V
\mathbbm{1}_{\{a_{j}^\text{L}+X_{-j}\geq \VaR_\alpha(X)\}}
]-\sum_{j\in I}b_{j}^\text{U}\bE[X_i V r_\psi(V)
\mathbbm{1}_{\{a_{j}^\text{U}+X_{-j}\geq \VaR_\alpha(X)\}}
],\\
B_{I}&=\sum_{j\in I}\frac{b_{j}^\text{L}}{\psi'(0)}\bE[V
\mathbbm{1}_{\{a_{j}^\text{L}+X_{-j}\geq \VaR_\alpha(X)\}}
]-\sum_{j\in I}b_{j}^\text{U}\bE[V r_\psi(V)
\mathbbm{1}_{\{a_{j}^\text{U}+X_{-j}\geq \VaR_\alpha(X)\}}
],\\
\pi_{iI}&\equiv 0\quad\text{and}\quad
\pi_{I}=\sum_{j\in I}
E'_{j}(X_j)-\gamma_{j}(\Ubold;\psi)f_{j}(X_j)
.
\end{align*}
For instance, the boundary terms of the survival Clayton and survival Gumbel copulas vanish when $b_j^\text{L}=0$ and $b_j^\text{U}=0$, respectively.

\subsection{Elliptical copula}\label{app:formulas:elliptical:copulas}

In this section we consider the case when the underlying copula is an elliptical copula arising from an elliptical distribution with density.
A $d$-dimensional random vector $\Z$ is said to follow an \emph{elliptical distribution} with mean vector $\mathbf{\mu}\in\bR^d$, dispersion matrix $\Sigma>0$ and a density generator $\xi_d:\bR_{+}\rightarrow \bR_{+}$, denoted by $\Y\sim \mathcal E_d(\mathbf{\mu},\Sigma,\xi_d)$, if $\Z$ admits a density of the form
\begin{align*}
f_{\Z}(\z)=\frac{r_d}{\sqrt{|\Sigma|}}\xi_d\left(\frac{1}{2}(\z-\mathbf{\mu})^{\top}\Sigma^\i(\z-\mathbf{\mu})\right),\quad \z \in \bR^d,
\end{align*}
where $r_d=r_d(\mathbf{\mu},\Sigma)$ is a normalization constant such that $\int_{\bR^d}f_{\Z}(\z) d\z=1$.
Let $P>0$ be a correlation matrix of $\Z$.
The copula of $\Z$ is called an \emph{elliptical copula}, and is denoted by $C_{P,\xi_d}$.
A $d$-dimensional random vector with marginal distributions $F_1,\dots,F_d$ and an elliptical copula $C_{P,\xi_d}$ can be modeled via
\begin{align}\label{eq:elliptical:model}
X_i=g_i(\Y)\quad\text{where}\quad g_i(\y)=F_i^\i(F_{Y_i}(y_i)),
\end{align}
where $\Y\sim \mathcal E_d(\mathbf{0}_d,P,\xi_d)$ and $F_{Y_i}$ is the cumulative distribution of $Y_i$.
We thus set $k=d$ and assume (MC1), (MC2) and that $\Y$ satisfies the following ``Elliptical Conditions":
\begin{enumerate}[leftmargin=*]
\item[(EC1)] $\xi_d\in {\mathcal C}^1(\bR_{+})$,\\
\item[(EC2)] $0<\xi_d(t)<\infty$ for $t\geq 0$, and\\
\item[(EC3)] $\lim_{y_j\rightarrow \infty}\frac{f_{\Y}(\y)}{f_{Y_j}(y_j)}=\lim_{y_j\rightarrow -\infty}\frac{f_{\Y}(\y)}{f_{Y_j}(y_j)}=0$ for every $\y_{-j}\in \bR^{d-1}$.
\end{enumerate}
We now derive the weights and boundary terms in Theorem \ref{thm:extension} under model \eqref{eq:elliptical:model}.

Since
\begin{align*}
\partial_{j}g_i(\y)=
\begin{cases}
\frac{f_{Y_i}(y_i)}{f_i(g_i(\y))}
,&\text{ if }j=i,\\
0, & \text{ if } j\neq i,
\end{cases}
\end{align*}
we take $j\neq i$, and then we have that
\begin{align*}
\partial_{j}g(\y)&=\sum_{m=1}^d \partial_{j}g_m(\y)=
\frac{f_{Y_{j}}(y_{j})}{f_{j}(g_{j}(\y))},
\end{align*}
which is non-zero a.e.\ by (EC2).
Next, we have, for $H_{ij}(\ubold) = g_i(\ubold)/\partial_j g(\ubold)$, that 
\begin{align*}
H_{ij}(\y)=\frac{g_i(\y)f_{j}(g_{j}(\y))f_{\Y}(\y)}{f_{Y_{j}}(y_{j})}.
\end{align*}
To study the boundary terms, we have that
\begin{align*}
\lim_{y_{j}\uparrow \infty}H_{ij}(\y)=
\lim_{y_{j}\downarrow -\infty}H_{ij}(\y)=0
\end{align*}
by (MC2) and (EC3).
Therefore, the boundary terms are all zero.
To compute the weight, we notice $\pi_{ij}\equiv 0$ by its definition.
We also find
\begin{align*}
\frac{\partial_{j}^2g(\y)}{(\partial_{j}g(\y))^2}&=\frac{f_{j}^2(g_{j}(\y))}{f_{Y_{j}}^2(y_{j})}\frac{f'_{Y_{j}}(y_{j})f_{j}(g_{j}(\y))-f_{Y_{j}}(y_{j}){f'}_{j}(g_{j}(\y))\frac{f_{Y_{j}}(y_{j})}{{f}_{j}(g_{j}(\y))}}{f_{j}^2(g_{j}(\y))}\\
&=-\frac{{f'}_{j}(g_{j}(\y))}{f_{j}(g_{j}(\y))}+\frac{{f'}_{Y_{j}}(y_{j})}{f_{Y_{j}}^2(y_{j})}f_{j}(g_{j}(\y)),
\end{align*}
and then
\begin{align*}
\frac{\partial_{j}f_{\Y}(\y)}{f_{\Y}(\y)}=\frac{{\xi'}_d\left(\frac{1}{2}\y^{\top}P^\i\y\right)}{\xi_d\left(\frac{1}{2}\y^{\top}P^\i\y\right)}\y^{\top}\mathbf{q}_{j},
\end{align*}
where $P^\i=Q=(\mathbf{q}_1,\dots,\mathbf{q}_d)$ for $\mathbf{q}_1,\dots,\mathbf{q}_d \in \bR^d$.
Therefore, we conclude
\begin{align*}
\pi_{j}(\y)&=
-\frac{{f'}_{j}(g_{j}(\y))}{f_{j}(g_{j}(\y))}+\frac{{f'}_{Y_{j}}(y_{j})}{f_{Y_{j}}^2(y_{j})}f_{j}(g_{j}(\y))
-\frac{{\xi'}_d\left(\frac{1}{2}\y^{\top}P^\i\y\right)}{\xi_d\left(\frac{1}{2}\y^{\top}P^\i\y\right)}\y^{\top}\mathbf{q}_{j}\frac{f_{j}(g_{j}(\y))}{f_{Y_{j}}(y_{j})}\\
&=E'_{j}(g_{j}(\y))+\gamma_{j}(\y;\xi_d,P)f_{j}(g_{j}(\y)),
\end{align*}
where
\begin{align*}
\gamma_{j}(\y;\xi_d,P)=\frac{{f'}_{Y_{j}}(y_{j})}{f_{Y_{j}}^2(y_{j})}-\frac{{\xi'}_d\left(\frac{1}{2}\y^{\top}P^\i\y\right)}{\xi_d\left(\frac{1}{2}\y^{\top}P^\i\y\right)}\frac{\y^{\top}\mathbf{q}_{j}}{f_{Y_{j}}(y_{j})}.
\end{align*}
Consequently, we derive the formula:
\begin{align*}
\mathscr{C}^{\alpha}_i=\frac{\bE\left[\pi_{I}X_i\,|\,X\geq \VaR_\alpha(X)\right]
}{\bE\left[\pi_{I}\,|\,X\geq \VaR_\alpha(X)\right]
},
\end{align*}
where
\begin{align*}
\pi_I = \sum_{j\in I} E'_{j}(X_j)+\gamma_{j}(\Y;\xi_d,P)f_{j}(X_j).
\end{align*}

\subsubsection{Gaussian copula}\label{example:gaussian:copulas}
Let $\Y\sim \Norm_d(\mathbf{0}_d,P)$ for a correlation matrix $P>0$.
Then (EC1) and (EC2) hold since $\xi_d(t)=\exp(-t)$, $t\geq 0$.
Moreover, we have that
\begin{align*}
\lim_{y_j\rightarrow \infty}\frac{f_{\Y}(\y)}{f_{Y_j}(y_j)}=\lim_{y_j\rightarrow \infty}\exp\left\{-\frac{1}{2}(\y^{\top}P^\i \y - y_j^2)\right\}=\lim_{y_j\rightarrow \infty}
\exp\left\{-\frac{1}{2}(q_{jj}-1)y_j^2\right\},
\end{align*}
where $q_{ij}$ is the $(i,j)$-th element of $P^\i$.
Since $q_{jj}\geq 1$, we have that $\lim_{y_j\rightarrow \infty}\frac{f_{\Y}(\y)}{f_{Y_j}(y_j)}=0$ if $q_{jj}\neq 1$.
Under this assumption, we similarly obtain $\lim_{y_j\rightarrow -\infty}\frac{f_{\Y}(\y)}{f_{Y_j}(y_j)} = 0$.
Therefore, (EC3) holds if $q_{jj}\neq 1$.
Since $\frac{f'_{Y_j}(y_j)}{f_{Y_j}^2(y_j)}=-\sqrt{2\pi}y_j\exp\left(\frac{1}{2}y_j^2\right)$ and $\frac{\xi_d'}{\xi_d}=-1$, we have that
\begin{align*}
\gamma_j(\y;\xi_d,P)&=-\sqrt{2\pi}y_j\exp\left(\frac{1}{2}y_j^2\right)+\sqrt{2\pi}\y^{\top}\mathbf{q}_j\exp\left(\frac{1}{2}y_j^2\right)\\
&=\sqrt{2\pi}(\y^{\top}\mathbf{q}_j-y_j)\exp\left(\frac{1}{2}y_j^2\right).
\end{align*}

\subsubsection{$t$ copula}\label{example:t:copulas}
Let $\Y\sim t(\mathbf{0}_d,P,\nu)$ for a correlation matrix $P>0$ and degrees of freedom $\nu>0$.
Then (EC1) and (EC2) hold since $\xi_d(t)=\left(1+\frac{2t}{\nu}\right)^{-\frac{d+\nu}{2}}$, $t\geq 0$.
Moreover, we find
\begin{align*}
\lim_{y_j\rightarrow \infty}\frac{f_{\Y}(\y)}{f_{Y_j}(y_j)}
=\lim_{y_j\rightarrow \infty}\frac{\left(1+\frac{\y^{\top}P^\i \y}{\nu}\right)^{-\frac{d+\nu}{2}}}{\left(1+\frac{y_j^2}{\nu}\right)^{-\frac{1+\nu}{2}}}
\propto\lim_{y_j\rightarrow \infty}y_j^{-(d-1)}=0,
\end{align*}
and that $\lim_{y_j\rightarrow -\infty}\frac{f_{\Y}(\y)}{f_{Y_j}(y_j)}=0$ in a similar manner.
Therefore, (EC3) is fulfilled.
Since 
\begin{align*}
\frac{f'_{Y_j}(y_j)}{f_{Y_j}^2(y_j)}=-\frac{\nu+1}{\nu}\frac{y_j}{r}\left(1+\frac{y_j^2}{\nu}\right)^{\frac{\nu-1}{2}},\quad
r=\frac{\Gamma\left(\frac{\nu+1}{2}\right)}{\sqrt{\nu\pi}\Gamma\left(\frac{\pi}{2}\right)},
\end{align*} 
and $\frac{\xi_d'(t)}{\xi_d(t)}=-\frac{d+\nu}{\nu}\left(1+\frac{2t}{\nu}\right)^{-1}$, we conclude
\begin{align*}
\gamma_j(\y;\xi_d,P)&=-\frac{\nu+1}{\nu}\frac{y_j}{r}\left(1+\frac{y_j^2}{\nu}\right)^{\frac{\nu-1}{2}}+\frac{d+\nu}{\nu}\left(1+\frac{\y^{\top}P^\i \y}{\nu}\right)^{-1}\frac{\y^{\top}\mathbf{q}_{j}}{r\left(1+\frac{y_j^2}{\nu}\right)^{-\frac{\nu+1}{2}}}\\
&=-\frac{\nu+1}{\nu}\frac{y_j}{r}\left(1+\frac{y_j^2}{\nu}\right)^{\frac{\nu-1}{2}}+\frac{d+\nu}{\nu}
\frac{\y^{\top}\mathbf{q}_{j}}{r}
\left(1+\frac{\y^{\top}P^\i \y}{\nu}\right)^{-1}\left(1+\frac{y_j^2}{\nu}\right)^{\frac{\nu+1}{2}}.
\end{align*}

\section{Numerical experiments}\label{sec:numerics}

\subsection{IBP estimators for parametric models}\label{subsec:parametric}

In order to assess the efficiency of an estimator derived from the IBP formulas \eqref{eq:risk_allocation_pi_uniform} and \eqref{eq:risk_allocation_pi_general} when compared to the $\delta$-allocations \eqref{eq:approx_contrib}, we perform simulation exercises for some of the models presented in Section \ref{sec:examples}.

To compare simple Monte Carlo estimators, we first pre-compute $\VaR_\alpha(X)$ and use it for both methods. The error in this step is irrelevant, as we can think we are computing $\Exp[X_i \mid X = B]$ where $B$ is a constant, so for the remainder of this section we assume all VaRs have been perfectly computed.
Let us denote by $A_{\alpha, \delta}$ the conditioning event in (\ref{eq:approx_contrib}), i.e. $A_{\alpha, \delta} = \{\VaR_{\alpha -\delta}(X)\leq X\leq   \VaR_{\alpha + \delta}(X)  \}$. We also define $A_\alpha = \{ X \geq \VaR_{\alpha}(X)\}$, the conditioning event in the IBP formulas \eqref{eq:risk_allocation_pi_uniform} and \eqref{eq:risk_allocation_pi_general}.

Let $\Y^{(1)}, \ldots, \Y^{(N)}$ be an i.i.d. sample of the driving random vector $\Y$ and let $\X^{(1)}, \ldots, \X^{(N)}$ be a sample of the random vector such that $X_i^{(n)}=g_i(\Y^{(n)})$ for $i=1,\dots,d$ and $n=1,\dots,N$.

We denote by 
\begin{align*}
N_\alpha^*=\sum_{n=1}^N \mathbbm{1}_{\{\X^{(n)}\in A_\alpha\}}\quad\text{and}\quad
N^*_{\alpha,\delta}=\sum_{n=1}^N \mathbbm{1}_{\{\X^{(n)}\in A_{\alpha,\delta}\}}.
\end{align*}

Note that $0 \leq N^*_\alpha, N_{\alpha,\delta}^* \leq N$, i.e., both $N_\alpha^*$ and $N^*_{\alpha,\delta}$ can be zero and in this case the estimates would not be defined. For a fixed sample size $N$, we have that  $\Exp[N^*_{\alpha}] = (1-\alpha)N$ and $\Exp[N^*_{\alpha,\delta}] = 2\delta N$, and thus we have $\Exp[N^*_{\alpha}] \geq \Exp[N^*_{\alpha,\delta}]$ as long as $1- \alpha \geq 2\delta$.

Based on \eqref{eq:approx_contrib}, we define the $\delta$-\emph{estimator} by
\begin{align}\label{eq:delta:estimator}
\widehat{\mathscr{C}}^{\alpha, \delta}_i=\frac{1}{N^*_{\alpha,\delta}}\sum_{n=1}^N X_i^{(n)}\mathbbm{1}_{\{\X^{(n)}\in A_{\alpha,\delta}\}}.
\end{align}
Similarly, based on the IBP formulas \eqref{eq:risk_allocation_pi_uniform} and \eqref{eq:risk_allocation_pi_general}, the \emph{IBP estimator} is defined by
\begin{align}\label{eq:IBP:estimator}
\widehat{\mathscr{C}}^\alpha_i &= 
\frac{
\widehat{B}_{iI}+(1-\alpha)\frac{1}{N_\alpha^*}\sum_{n=1}^N\{X_i^{(n)}\pi_I(\Y^{(n)})+\pi_{iI}(\Y^{(n)})\}\mathbbm{1}_{\{\X^{(n)}\in A_{\alpha}\}}
}{
\widehat{B}_{I}+(1-\alpha)\frac{1}{N_\alpha^*}\sum_{n=1}^N\pi_I(\Y^{(n)})\mathbbm{1}_{\{\X^{(n)}\in A_{\alpha}\}}
},
\end{align}
where the weights in \eqref{eq:weights_general} are regarded as functions of $\Y$, and $\widehat{B}_{iI}$ and $\widehat{B}_{I}$ are respective estimators of $B_{iI}$ and $B_I$ determined based on the underlying model. 
For copula models listed in Table \ref{tab:weights:copulas}, $\widehat{B}_{iI}$ and $\widehat{B}_{I}$ are linear combinations of some expectations.
Therefore, we determine $\widehat{B}_{iI}$ and $\widehat{B}_{I}$ to be the corresponding linear combinations of sample means. 
Consequently, the $\delta$-estimator \eqref{eq:delta:estimator} and the IBP estimator \eqref{eq:IBP:estimator} are consistent estimators of \eqref{eq:approx_contrib} and \eqref{eq:var_allocation}, respectively.

However, as an estimator of VaR contribution \eqref{eq:var_allocation}, the $\delta$-estimator \eqref{eq:delta:estimator} is not consistent since the bias term $\mathscr{C}_i^{\alpha,\delta}-\mathscr{C}_i^\alpha$ does not vanish as $N \rightarrow \infty$.
In fact, there is a trade-off relation between $\delta$ and $N$ in terms of the bias and the standard error of the $\delta$-estimator; see \cite[Section~2]{koike2019estimation} for details.
To study the impact of $\delta$ and the sample size $N$ in the mean and variance of both estimators, we compute the $10\,000$ replicates of the estimators \eqref{eq:delta:estimator} and \eqref{eq:IBP:estimator}  for $\alpha \in \{0.5, 0.9, 0.99\}$ for each combination of $\delta \in \{10^{-3}, 10^{-4}, 10^{-5}, 10^{-6}\}$ and $N \in \{10^4, \lfloor 10^{4.5}\rfloor, 10^5, \lfloor10^{5.5}\rfloor, 10^6\}$. 
Although the case $\alpha = 0.5$ may not be of primary concern in risk management practice, we expect the IBP estimator to be numerically more stable than the $\delta$-estimator for smaller values of $\alpha$ since more samples fall in the conditioning event $A_\alpha$. 
Lower variance of the IBP estimator is also expected compared with the $\delta$-estimator since the effective sample size $\Exp[N^*_{\alpha}] = (1-\alpha)N$ of the IBP estimator is typically larger than that of the $\delta$-estimator $\Exp[N^*_{\alpha,\delta}] = 2\delta N$.
Note that all comparisons are naturally adjusted by their computational cost, as both estimators use the same number of samples. 
In fact, both estimators are computed based on exactly the same sample.
As the results are qualitatively similar across the components of $\X$, for each model we only present the results for its first component, i.e. $\mathscr{C}_1^\alpha$.

\subsubsection{Multivariate Gaussian} \label{sec:euler_multi_gauss}
If we assume the vector of risks follows a multivariate normal distribution, i.e., $\X = (X_1,\hdots,X_d) \sim N(\mathbf{0}, \Sigma)$ for some positive definite covariance matrix $\Sigma$, then the total VaR and the VaR contributions can be computed in closed form, as
\begin{align}\label{eq:formula:gauss}
\VaR_{\alpha}(X) = \Phi^{-1}(\alpha)\sqrt{\mathbf{1}^T\Sigma\mathbf{1}}
\quad\text{and}\quad
\mathscr{C}^{\alpha}_i = 
 \frac{(\Sigma \mathbf{1})^T_i}{\mathbf{1}^T\Sigma\mathbf{1}}\,\VaR_{\alpha}(X),
\end{align}
respectively; see \cite[Corollary~8.43]{mcneil2015quantitative}. 
In our example we take $d=3$ and $\Sigma = L L^T$, with
$$
L = \begin{pmatrix}
1 & 0 & 0 \\ 0.5 & 0.7 & 0 \\ 1 & 0.8 & 1.1
\end{pmatrix}\quad\text{and}\quad
\Sigma = \begin{pmatrix}
1 & 0.50 & 1 \\ 0.50 & 0.74 & 1.06 \\ 1 & 1.06 & 2.85
\end{pmatrix}.
$$ 

\begin{figure}
	\includegraphics[width=\linewidth]{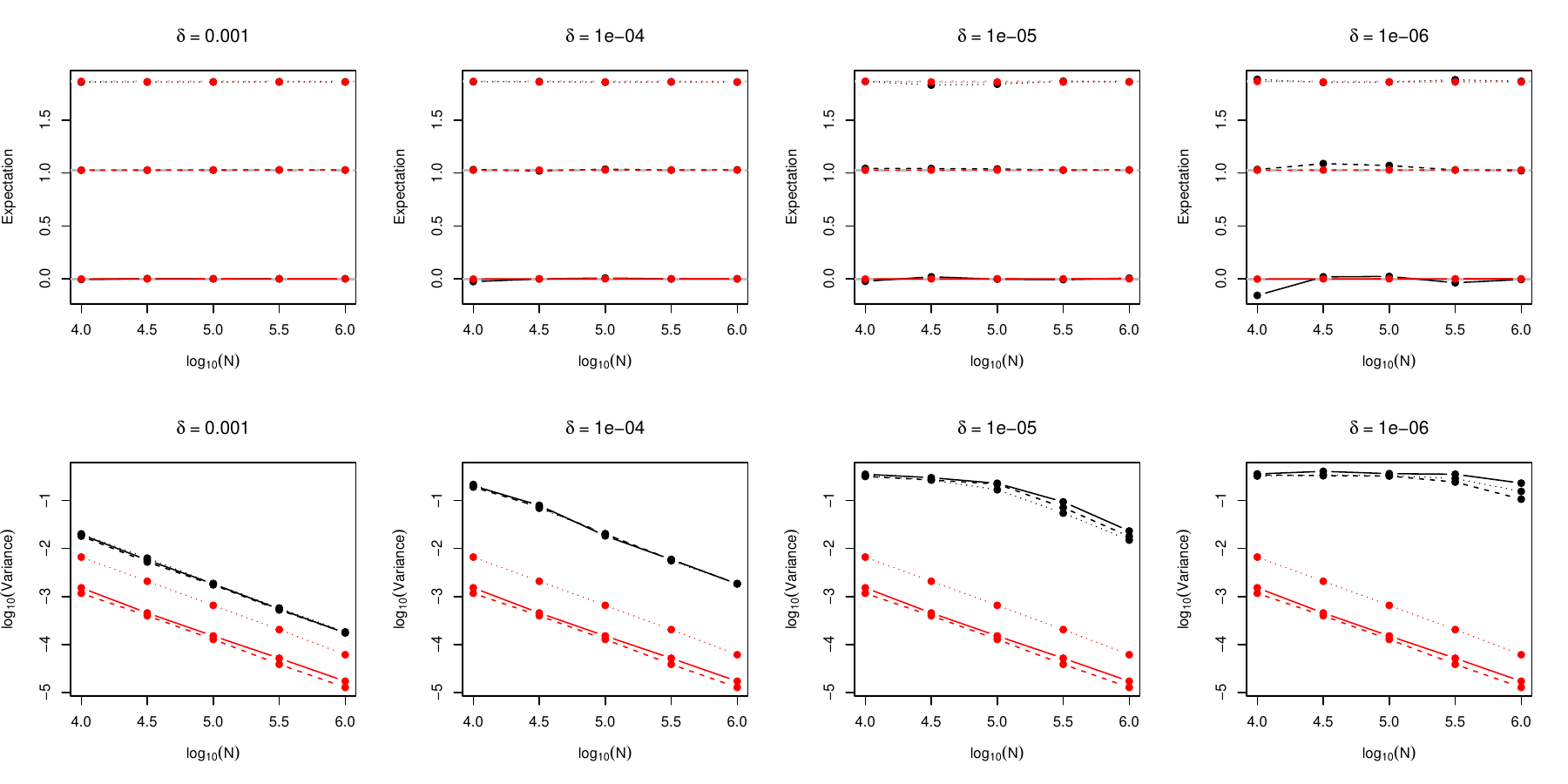}
	\label{fig:Gaussian}
	\caption{Multivariate Gaussian model of Section \ref{sec:euler_multi_gauss}. Mean (top row) and variance (bottom row) of the $\delta$-estimator (black) and the IBP (red) for the first allocation $\mathscr{C}_1^\alpha$. Each line type (solid, dashed and dotted) represent a different value of $\alpha\in \{0.5,0.9,0.99\}$. The columns represent different values for $\delta$. Solid gray lines in the Expectation plots denote the true allocations.}
\end{figure}

From Figure \ref{fig:Gaussian} we see that both estimators are unbiasedly reproducing the allocations calculated from the closed form expression (gray lines) for all values of $\alpha$. When the value of $\delta$ is too small ($\delta = 10^{-6}$), we see the $\delta$-estimator needs more samples to reduce the bias.

The striking result is related to the variance. As expected, when $\delta$ gets smaller the variance of the $\delta$-estimator (for the same sample size) depletes very quickly since there are less samples belonging to $A_{\alpha,\delta}$. On the other hand, for the IBP estimator the proportion of samples belonging to $A_\alpha$ is on average $1-\alpha$, whilst it is $2\delta$ for the $\delta$-allocation. It can also be seen from Figure \ref{fig:Gaussian} that, although all variances converge exponentially fast to zero as $N$ increases, the IBP variances are consistently smaller than the $\delta$-estimator. Moreover, the IBP estimator does not depend on the $\delta$ parameter, which, if chosen incorrectly in the $\delta$-estimator, may hinder the variance convergence.

\subsubsection{Independent log-normals} \label{sec:indep_LN}

\begin{figure}
	\includegraphics[width=\linewidth]{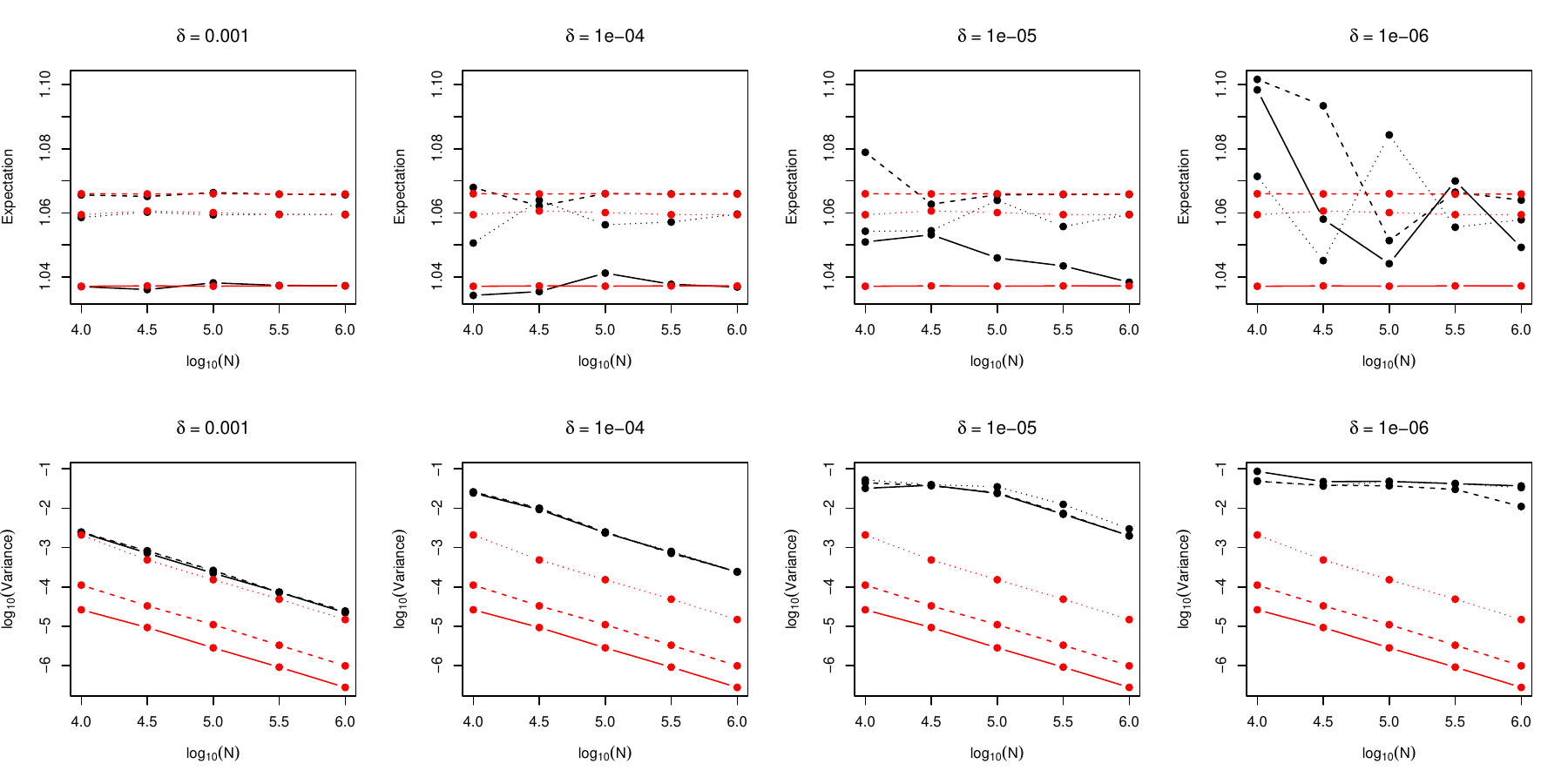}
	\label{fig:indep_LN}
	\caption{Independent Log-Normals model of Section \ref{sec:indep_LN}. Mean (top row) and variance (bottom row) of the $\delta$-estimator (black) and the IBP (red) for the first allocation $\mathscr{C}_1^\alpha$. Each line type (solid, dashed and dotted) represent a different value of $\alpha\in \{0.5,0.9,0.99\}$. The columns represent different values for $\delta$.}
\end{figure}

For this example we follow the independence structure of Section \ref{sec:ex_indep} and assume $X_i \stackrel{ind}{\sim} LN(0,\sigma_i)$ with $\sigma_1 = 0.2$, $\sigma_2 = 0.7$ and $\sigma_3 = 0.5$. Note that in this case the boundary terms $B_{iI}$ and $B_I$ vanish; see Table \ref{tab:weights:margins}. The results are presented in Figure \ref{fig:indep_LN}.

In this seemingly simple example where no closed form expression is available, we can clearly see the impact of the choice of $\delta$ in the $\delta$-estimator. Even though for the largest $\delta$ under consideration the $\delta$-estimators are reasonably precise, when $\delta$ decreases the $\delta$-estimator presents considerable bias for small $N$. When $N$ increases, though, the estimates converge to the IBP ones (apart from the 50\% quantile when $\delta = 10^{-6}$). The estimators' variances behave very similarly to the Gaussian case, with the IBP estimator performing better in all scenarios.

\subsubsection{Independent skew $t$}\label{sec:skew_t}
\begin{figure}
	\includegraphics[width=\linewidth]{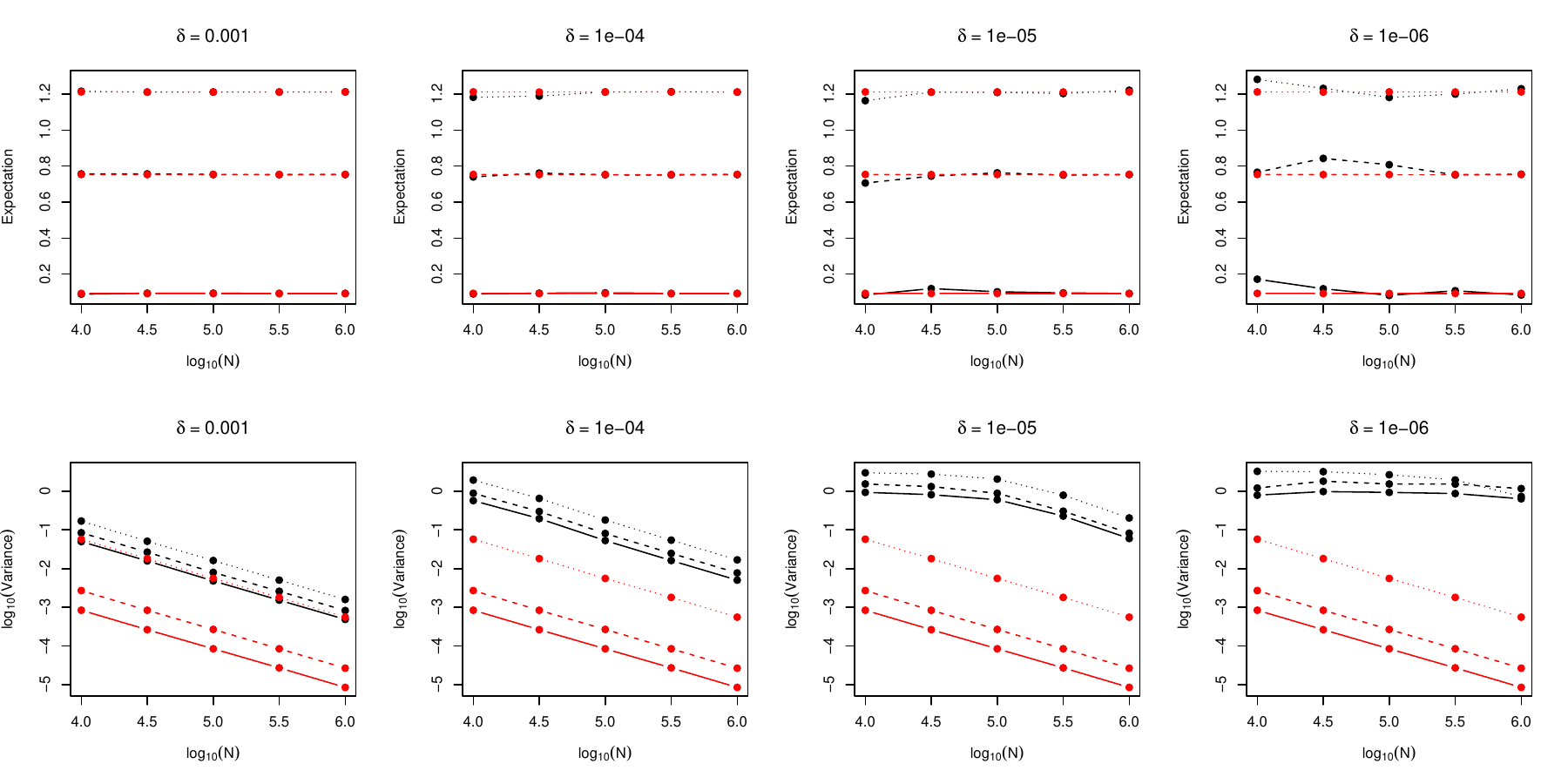}
	\label{fig:indep_skt}
	\caption{Independent skew $t$ model from Section \ref{sec:skew_t}. Mean (top row) and variance (bottom row) of the $\delta$-estimator (black) and the IBP (red) for the first allocation $\mathscr{C}_1^\alpha$. Each line type (solid, dashed and dotted) represent a different value of $\alpha\in \{0.5,0.9,0.99\}$. The columns represent different values for $\delta$.}
\end{figure}

In order to better understand the impact of skewness in the IBP estimator, we study and example where $X_i \stackrel{ind}{\sim} Skt(\nu_j, \gamma_j)$ (using the notation from \cite{fernandez1998bayesian}), with $\nu_1 = 5, \nu_2 = 5.5, \nu_3 = 6$ and $\gamma_1 = 1, \gamma_2 = 1.5, \gamma_3 = 2$. In this case the boundary terms $B_{i,I}$ and $B_I$ also vanish (see Table \ref{tab:weights:margins}). Apart from the fact that the $\delta$-estimator is less biased in this example when compared to the Independent Log-Normals from Section \ref{sec:skew_t}, the qualitative behavior seen in Figure \ref{fig:indep_skt} is very similar to Figure \ref{fig:indep_LN}.	

\subsubsection{Clayton copula with Gaussian marginals} \label{sec:clayton_gaussian}
This simulation exercise is related to Section \ref{sec:ex_clayton}, where the dependence structure of the multivariate vector $\X$ is assumed to be given by a Clayton copula. Here we assume the copula parameter is $\vartheta = 2$ and the marginals are $X_i \sim N(0, \sigma^2_i)$, with $\sigma_1 = 1$, $\sigma_2 = 0.5$ and $\sigma_3 = 1$. Due to the Gaussian marginals, the boundary terms vanish in this case.

\begin{figure}
	\includegraphics[width=\linewidth]{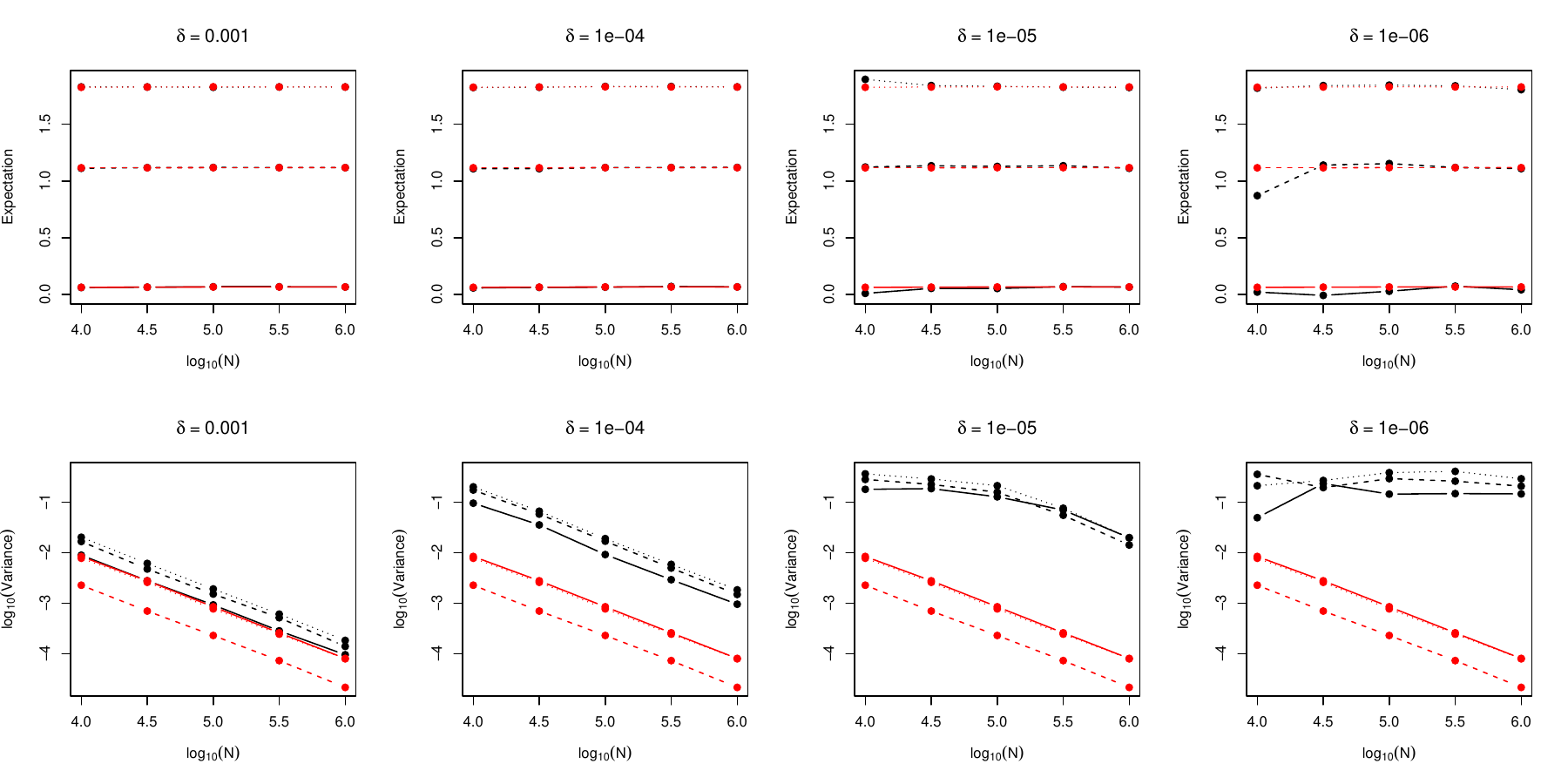}
	\label{fig:clayton_gaussian2}
	\caption{Clayton copula with Gaussian marginals of model in Section \ref{sec:clayton_gaussian}. Mean (top row) and variance (bottom row) of the $\delta$-estimator (black) and the IBP (red) for the first allocation $\mathscr{C}_1^\alpha$. Each line type (solid, dashed and dotted) represent a different value of $\alpha\in \{0.5,0.9,0.99\}$. The columns represent different values for $\delta$.}
\end{figure}

The mean and variance of the estimators are presented in Figure \ref{fig:clayton_gaussian2}. Once again the results are qualitatively equivalent to the others, with the additional feature that for $\delta = 10^{-6}$ there are convergence issues for the $\delta$-estimator when the sample size $N$ is smaller than $10^{5}$. For the variance, the IBP estimator performs uniformly better in all scenarios.

\subsubsection{Survival Clayton copula with GPD marginals} 
\begin{figure}
	\includegraphics[width=\linewidth]{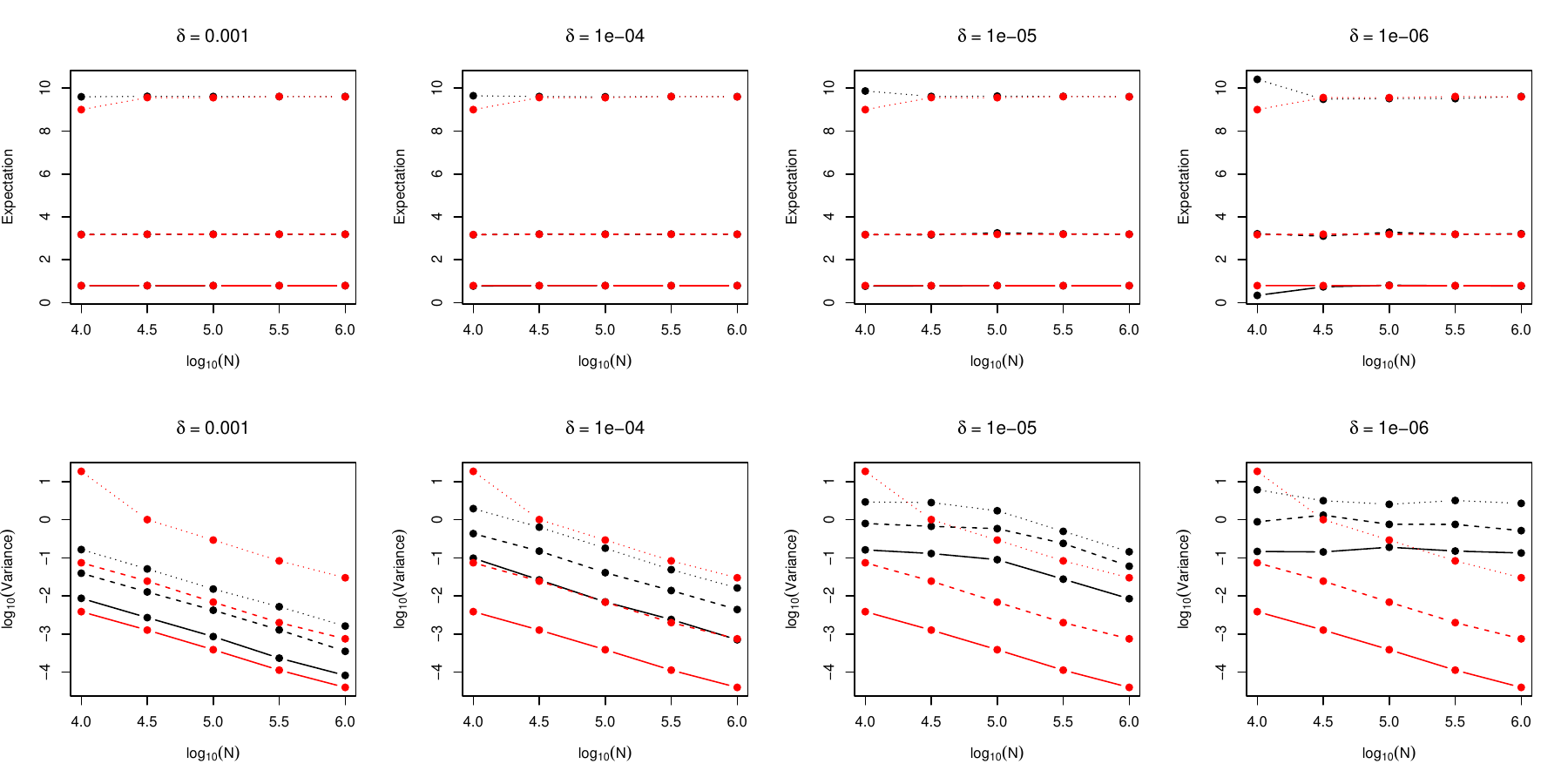}
	\label{fig:survivalClayton_GPD}
	\caption{Survival Clayton copula with GPD marginals. Mean (top row) and variance (bottom row) of the $\delta$-estimator (black) and the IBP (red) for the first allocation $\mathscr{C}_1^\alpha$. Each line type (solid, dashed and dotted) represent a different value of $\alpha\in \{0.5,0.9,0.99\}$. The columns represent different values for $\delta$.}
\end{figure}

For the last computational experiment, we use the model (M1) from \cite{koike2020markov}, where it is assumed $X_i \sim GPD(\xi_i, \beta_j)$ with $\xi_i = 0.3$ and $\beta_i = 1$, for $i=1,2,3$. The results are found in Figure \ref{fig:survivalClayton_GPD}. 

As with all other examples, we see the IBP estimators do not show any bias for large enough sample size, i.e., $N> 10^4$, whereas the $\delta$-estimator seems to be biased for the 99\% and 50\% quantiles when $\delta = 10^{-6}$.

The decay of the variance of the $\delta$-estimator with respect to the sample size $N$ is strongly dependent on the value of $\delta$, to the point that no convergence is observed when $\delta = 10^{-6}$. Nevertheless, for large enough values of $\delta$ the $\delta$-estimator's variance happens to be smaller than the IBP's. This is the case when $\delta = 10^{-3}$ and $\alpha\in \{0.9,0.99\}$ for all values of $N$.

For $\delta = 10^{-4}$, the IBP estimator is more volatile than the $\delta$-estimator only when $\alpha=0.99$. For smaller values of $\delta$ the IBP estimator presents lower variance except when $N=10^4$.

\subsection{Nonparametric IBP estimator with kernel density estimation}\label{subsec:kde}

Even though the focus of this paper is not on the full workflow described in the Introduction, here we briefly discuss how the IBP formulation of the VaR contributions can be used for estimation of the contributions under a nonparametric loss model.

We assume that after observing $M$ losses denoted by $\{\x_1,\ldots,\x_M\}$, the user decides to follow a nonparametric route, assuming the loss distribution is well approximated by a mixture of Gaussian densities, as in a Kernel Density Estimator (KDE). As per the discussion in the Introduction, we assume the items (a)-(d) have been previously performed and focus only on item (e): computing the VaR contributions when the underlying model is given by a KDE. To do so, we first need to pre-compute the total VaR and for this, we assume the underlying model is the KDE and compute the VaR based on this model using a large simulation, as in the previous examples. This VaR is used both in the $\delta$ and IBP allocations under the KDE model assumption.

The KDE of the underlying density of the losses in this case can be written as
$$\hat{f}(\x) = \frac{1}{M} \sum_{m=1}^M |\bfH|^{-1/2} K(\bfH^{-1/2}(\x - \x_m)),$$
where $\bfH$ is the bandwidth matrix (symmetric and positive definite), $|\bfH|$ its determinant and $K$ is a kernel function here assumed to be Gaussian. See \cite{silverman2018density} for a detailed exposition on kernel density estimators. Since this KDE is a mixture of normal distributions, the corresponding weights and boundary terms of the IBP formula are computed as in Example \ref{example:mixture}.

For the numerical example, we generate data $\{\x_1,\ldots,\x_M\}$, $M=100$, independently from $N(\mathbf{0}, \Sigma)$, where $\Sigma$ is as specified in Section \ref{sec:euler_multi_gauss}. The bandwidth matrix is chosen as in \cite[Equation (3.2)]{chacon2011asymptotics}, with $r=1$:
$$\widehat{\bfH} = \left[\frac{4}{M(d+2r+2)}\right]^{2/(d+2r+4)} \widehat{\Sigma}\quad\text{with}\quad
\widehat{\Sigma} = \begin{pmatrix}
1.09 & \phantom{-}0.42 & \phantom{-}0.86 \\ 
0.42 & \phantom{-}0.57 & \phantom{-}0.82 \\ 
0.86 & \phantom{-}0.82 & \phantom{-}2.24 \\ 
\end{pmatrix},
$$
where  $\widehat{\Sigma}$ is the empirical covariance matrix of the observed data $\{\x_1,\ldots,\x_M\}$. Note that this is also the choice in the function \texttt{Hns} of the \texttt{R} package \texttt{ks} \cite{ksPackage}.

\begin{figure}
	\includegraphics[width=\linewidth]{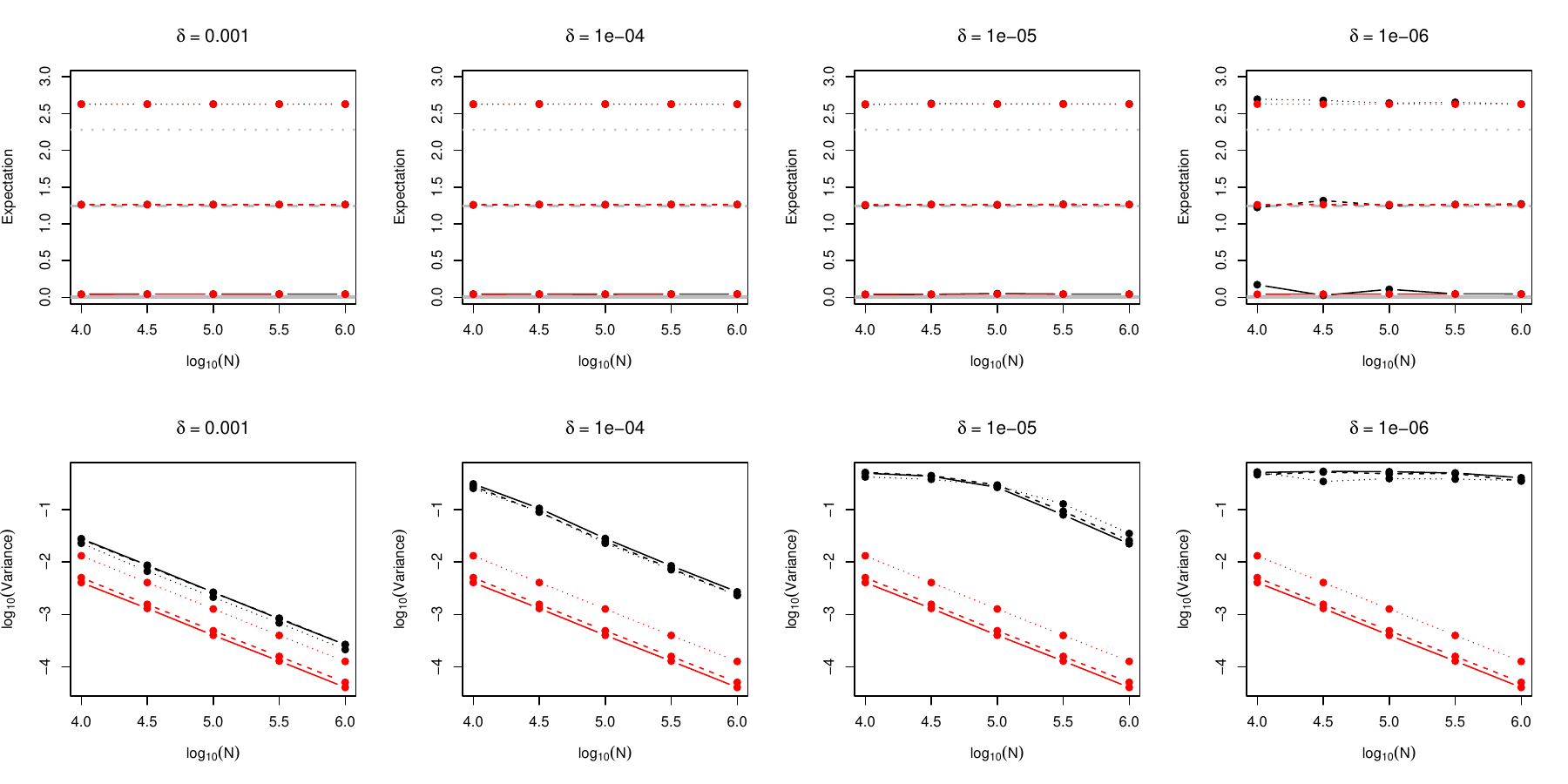}
	\caption{Nonparametric model from Section \ref{subsec:kde}. Mean (top row) and variance (bottom row) of the $\delta$-estimator (black) and the IBP (red) for the first allocation $\mathscr{C}_1^\alpha$. Each line type (solid, dashed and dotted) represent a different value of $\alpha\in \{0.5,0.9,0.99\}$. The columns represent different values for $\delta$. Gray lines in the Expectation plots denote allocations under the Gaussian model assumption (see Section \ref{subsec:kde} for details).}
	\label{fig:kde}
\end{figure}

\begin{figure}
	\includegraphics[width=\linewidth]{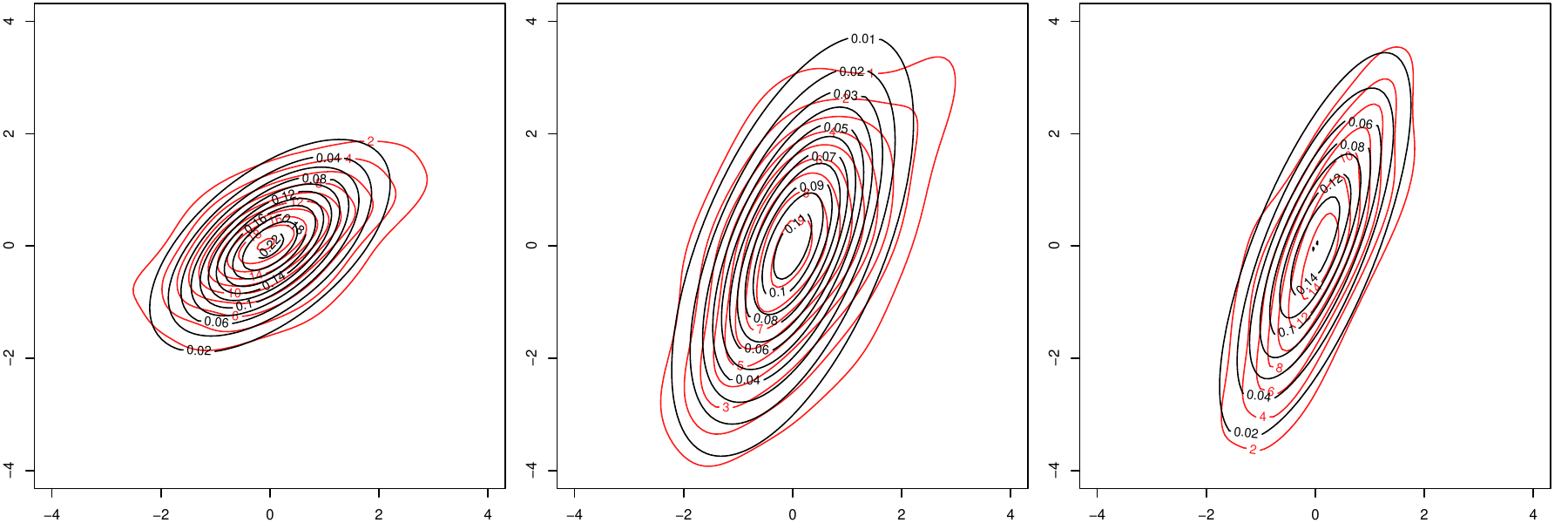}
	\caption{ Level sets for the true Gaussian distribution (black) and the KDE approximation (red) from Section \ref{subsec:kde}. From left to right, we have the pairs of marginals (1,2), (1,3) and (2,3).}
	\label{fig:kde_levels}
\end{figure}

From Figure \ref{fig:kde} we do not see any convergence issues for the estimators, apart from the delta allocation with $N < 10^6$ and $\delta = 10^{-6}$. The behavior of the variances is similar to the previous examples, with the IBP estimator been uniformly less volatile than the $\delta$ estimator, with variance decaying exponentially fast in the simulated sample size $N$. In order to provide some ground for comparison, the gray lines in Figure \ref{fig:kde} represent the estimates of VaR contributions under the Gaussian assumption, that is, the pre-computed total VaR is allocated in proportion to $(\widehat\Sigma \mathbf{1})^T$; see \eqref{eq:formula:gauss}.
The gray lines slightly deviate from the other lines for the case of high confidence level $\alpha=0.99$, which indicate the limitation of the KDE approximation to the original Gaussian distribution especially in the tail, see Figure \ref{fig:kde_levels}. 
In summary, although the choice of nonparametric method remains an issue, the IBP estimator is a more stable estimator than the $\delta$-estimator. 

\subsection{Real data analysis}\label{real:data:analysis}
In this section we compare nonparametric estimators of VaR contributions computed from a real data set.
We analyze daily logarithmic returns of NIKKEI and FTSE stock indices from 2005 to 2015 (10 years). 
The sample size is $T=2\,699$ days and the dimension is $d=2$.

Following Section~5.2 of~\cite{koike2019estimation}, GARCH(1,1) models with skew-$t$ white noise are fitted to marginal return series. 
Their dependence structure is induced by the copula, denoted by $C$, between the marginal white noises; see Sections~4 and~14 in~\cite{mcneil2015quantitative} for details of this model.
Differently from \cite{koike2019estimation}, which assumes a parametric copula, here no parametric model is assumed on the copula $C$.
Therefore, these model assumptions can be seen as semi-parametric, as the marginals are given by parametric models and the copula by a non-parametric one.

\begin{figure}
\begin{center}
	\includegraphics[width=0.5\linewidth]{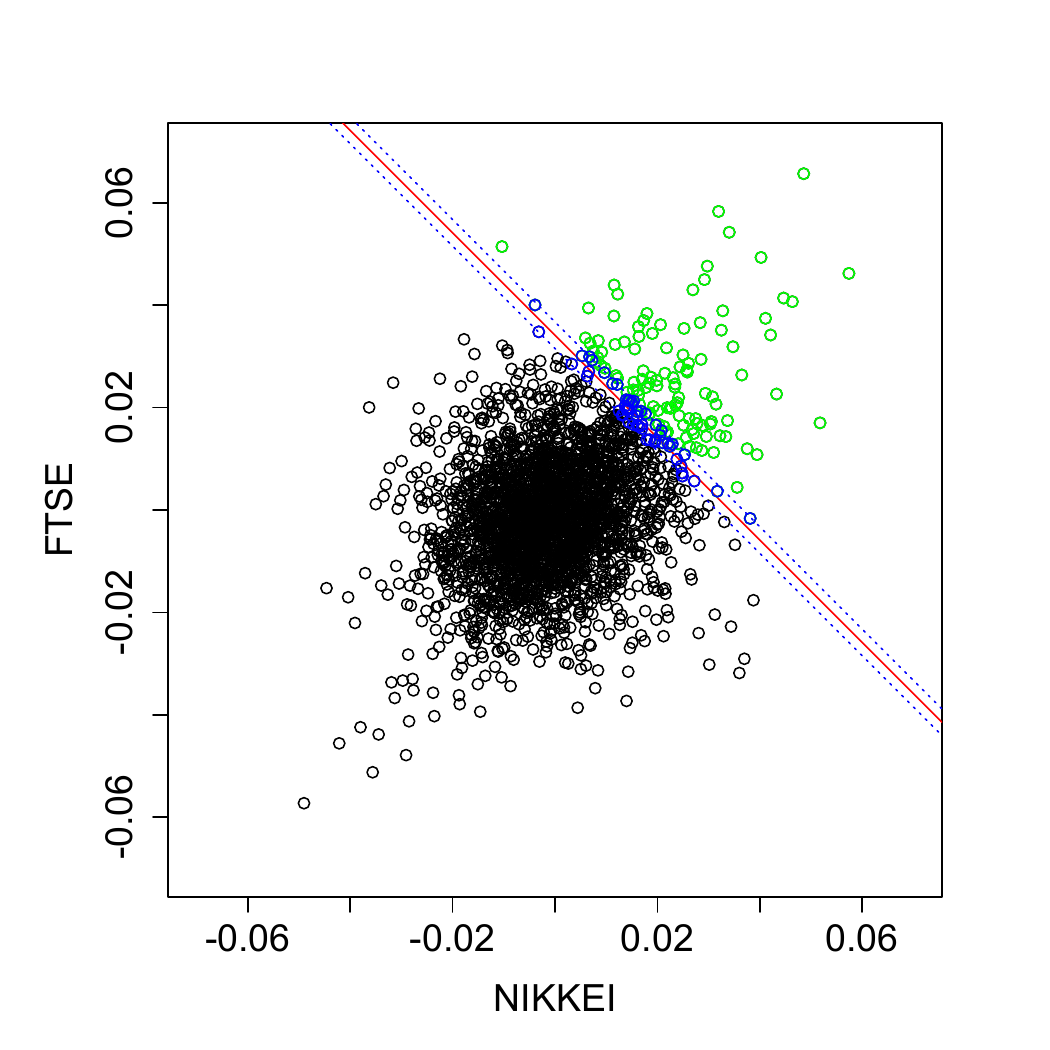}
	\end{center}
	\caption{Scatter plot of the joint returns $(X_{T+1,1}, X_{T+1,2})$ of NIKKEI and FTSE at time $T+1$ predicted from the observations up to time $T$.
	To estimate VaR contributions, the $\delta$-estimator takes average over samples highlighted in blue (around the red line) whereas the IBP estimator is obtained as a weighted average over samples highlighted in green (above the red line).}
	\label{fig:dataplot}
\end{figure}

The goal of this analysis is to compute the conditional VaR contributions of returns at time $T+1$ given all observations up to time $T$, denoted as $\mathcal{F}_T$.
The confidence level is fixed to be $\alpha = 0.95$ in this study.
Under the model described above, the marginal distributions of $X_{T+1,i} \mid \mathcal F_T$, the $i$th return at time $T+1$, for $i=1,2$, follow skew $t$ distributions with parameters induced by the marginal GARCH models.
Moreover, the copula of the joint return $\X = (X_{T+1,1}, X_{T+1,2}) \mid \mathcal{F}_T$ equals $C$, that of the white noises. 
Therefore, under this model assumption where marginal time series are parametrically specified while the model of the copula remains nonparametric, iid samples from $\X$ are extracted from data and we are able to estimate its VaR contributions nonparametrically based on this iid sample.
The scatter plot of this predicted joint returns at time $T+1$ is given in Figure~\ref{fig:dataplot}.

In order to compute VaR contributions with the proposed IBP method, we consider the trivial model in Remark \ref{example:trivial:driver}, where the driver is set to be $\Y = \X$ and the weight functions are given by the score function of $\X$. Since the returns take values in $\R$ it is reasonable to assume all boundary terms vanish in Theorem \ref{thm:extension}. Therefore, one only needs to compute the score function of $\X$. As the data generating model is assumed unknown, the score function needs to be estimated from observed data. This problem has been thoroughly discussed in the literature, e.g. ~\cite{strathmann2015gradient},~\cite{li2017gradient},~\cite{sriperumbudur2017density},~\cite{shi2018spectral},~\cite{sutherland2018efficient},~\cite{song2019generative} and~\cite{zhou2020nonparametric}.

We use a plug-in estimator to estimate the score function of $\X$ as considered in~\cite{li2017gradient}.
As seen in Remark \ref{rmk:score}, the IBP weights can be computed as a score function of $\X \mid \{X \geq \VaR_{\alpha}(X)\}$.
The plug-in estimator of the score function of $\X$ is then given by replacing the conditional density $f_{\X \mid \{X \geq \VaR_{\alpha}(X)\}}(\x)$ in the score function with its kernel density estimator based on the i.i.d. samples from $\X \mid \{ X \geq \VaR_{\alpha}(X)\}$. Differently from the previous exposition in Section~\ref{subsec:kde}, the KDE approximation is performed only in the region $\{\x \in \R^d: x_1+\cdots + x_d \geq \VaR_{\alpha}(X)\}$ of interest. 
This procedure can be followed if the only interest is on computing VaR contributions, whereas the one presented in Section~\ref{subsec:kde} would be preferred if the analyst plans to use the KDE approximation for other ends. 
Although other kernels are possible to choose, we consider a Gaussian kernel as in Section \ref{subsec:kde} since various methods for bandwidth selection are accessible.

The ingredients of the IBP estimator are completed with the choice of an index set $I$, which is taken as $I=\{2\}$ to compute $\mathscr{\hat C}_1^\alpha$ and  $I=\{1\}$ to compute $\mathscr{\hat C}_2^\alpha$.

To investigate sensitivity to the estimators and their hyperparameters, we compare the following seven estimators of VaR contributions;

\begin{enumerate}
\item[] (E1)--(E2): the $\delta$-estimators with $\delta = 0.005$ and $0.01$, respectively;
\item[] (E3)--(E7): the IBP estimators with the bandwidth matrices determined by the functions \texttt{Hns}, \texttt{Hpi}, \texttt{Hpi.diag}, \texttt{Hscv} and \texttt{Hscv.diag} in the \texttt{R} package \texttt{ks}, respectively.
\end{enumerate}

The total VaR estimated from the original data is $3.413\%$.
Therefore, $1\%$ of samples whose sums are around the total VaR are extracted for estimation of VaR contributions in (E1), and  $2\%$ of samples are used in (E2).
For (E3)--(E7), different methods are used to determine the bandwidth matrices; see \texttt{R} package \texttt{ks} \cite{ksPackage} and the references therein.
Note, in particular, that \texttt{Hpi.diag} and \texttt{Hscv.diag} choose diagonal bandwidth matrices.

\begin{figure}
	\includegraphics[width=\linewidth]{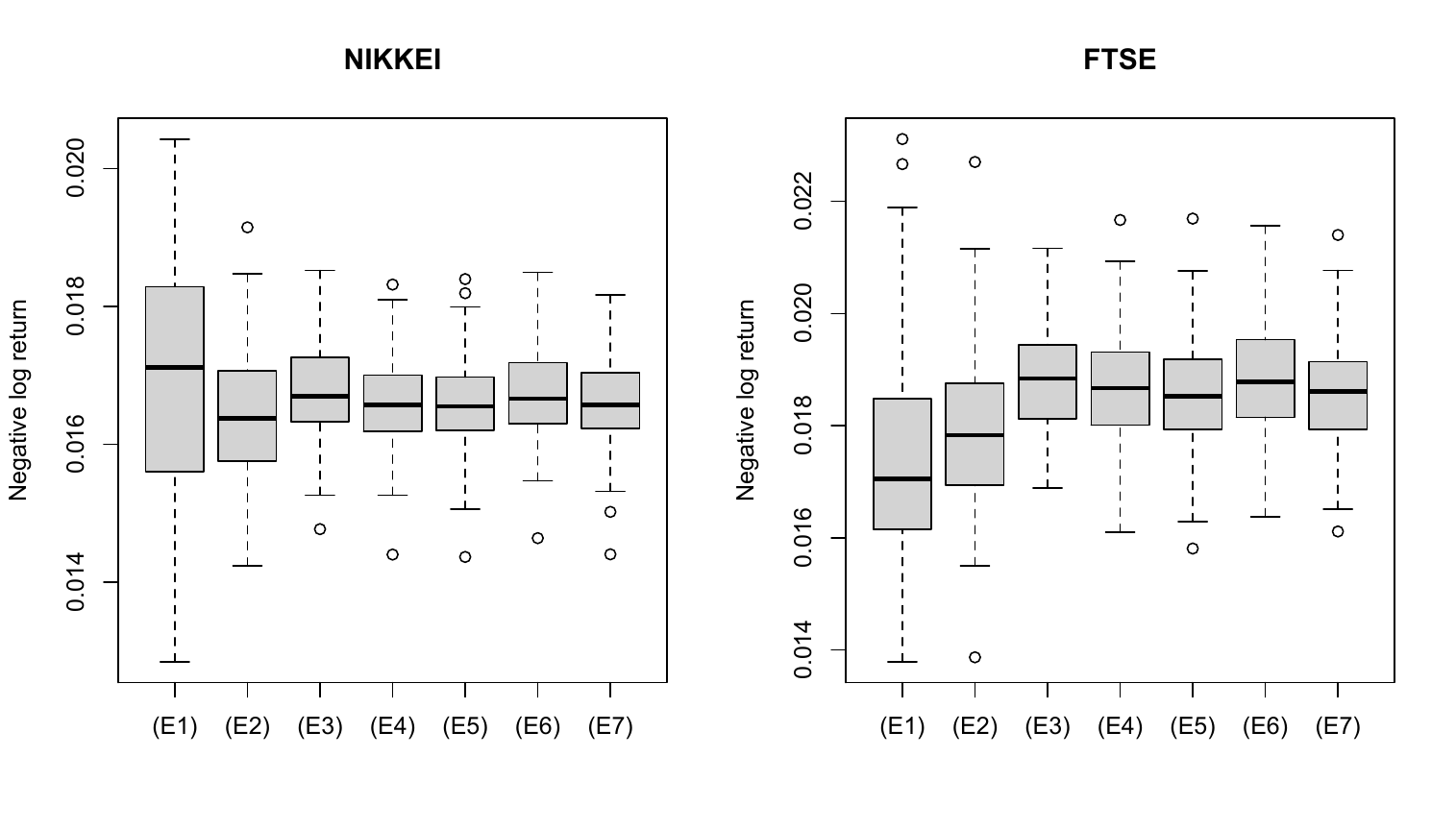}
	\caption{ Boxplots of $B=100$ bootstrap replicates of the estimators (E1)--(E7) of VaR contributions of DJ (left) and FTSE (right).}
	\label{fig:boxplots}
\end{figure}

For the estimators (E1)--(E7), we conduct a bootstrap analysis.
We compute $B=100$ bootstrap replicates of the estimators (E1)--(E7) where each estimate is computed from bootstrapped samples with size $T$ resampled from the original samples with replacement.
Figure~\ref{fig:boxplots} shows the boxplots of the bootstrap replicates of each estimator (E1)--(E7) to compare the variability of estimates.

From Figure~\ref{fig:boxplots}, we first observe that estimates of all the estimators are fairly closed with each other, that is, all the different estimators in this experiment provide robust estimates of VaR contributions.
On the other hand, IBP estimators (E3)--(E7) tend to have lower variability than the $\delta$-estimators (E1)--(E2).
In particular, IBP estimators with different bandwidth matrices provide quite similar results.
Therefore, the standard choice of Gaussian kernel provides stable results of the IBP estimator although other kernels may improve the performance.
In summary, for the case of real data analysis where the true model is unknown, the nonparametric IBP estimator may provide accurate and stable estimate of VaR contributions.

\section{ Conclusion and discussion } \label{sec:conclusion}

We are able to derive a novel expression for the Value-at-Risk contributions called the  \emph{IBP formula} in the form of a ratio of expectations (with some boundary terms) conditional on events of positive probability.
Since we avoid an expectation conditional on  a zero probability event in the usual representation, the new formulation is amenable to Monte Carlo simulation with mild hypothesis on the multivariate models.
Analytic formulas are also provided for a wide range of models.
The so-called \emph{IBP estimator} is then derived based on the IBP formula, and its performance is demonstrated in a series of simulation studies and a real data analysis.
We show that the IBP estimator outperforms the standard Monte Carlo estimator in various settings.

We end this paper with some potential avenues for future research.
First, as the expectations in the proposed formulation resemble the ES contributions, one could easily adapt the sampling schemes from, for example, \cite{targino2015sequential} and \cite{koike2020markov} for further computational gains.
Another relevant avenue for further improvements is to explore better estimators for the ratio of expectations. The methods in \cite{iglehart1975simulating} have been tested but it did not result in substantial bias reduction. 

Second, further investigation of the IBP estimator in the nonparametric models is required, such as consistency, asymptotic normality, kernel and its bandwidth selections.
Third, sensitivity of the IBP estimator to model misspecification is worth exploring. We remark that the IBP estimator could be sensitive to the underlying model since the reformulation from \eqref{eq:var_allocation} to \eqref{eq:var_allocation_intro} relies on the model structure, especially on the upper tail.
Finally, the methodology described in this paper can be applied to non-linear portfolio and to compute quantile sensitivities, as per Theorem 2 of \cite{hong2009estimating}.

\section*{Declaration of competing interest}
The authors declare that they have no known competing financial interests or personal relationships that could have appeared to influence the work reported in this paper.

\section*{Acknowledgements}
We are grateful to the handling editor and two anonymous referees for their careful reading of the manuscript and their insightful comments. We would also like to thank Ruodu Wang and Andreas Tsanakas for constructive comments on an earlier version of the paper.

\section*{Funding}
Takaaki Koike was supported by JSPS KAKENHI Grant Number JP21K13275.
%
 \bibliographystyle{spmpsci}
 \bibliography{bibliography}

\begin{thebibliography}{10}
\providecommand{\url}[1]{{#1}}
\providecommand{\urlprefix}{URL }
\expandafter\ifx\csname urlstyle\endcsname\relax
  \providecommand{\doi}[1]{DOI~\discretionary{}{}{}#1}\else
  \providecommand{\doi}{DOI~\discretionary{}{}{}\begingroup
  \urlstyle{rm}\Url}\fi

\bibitem{solvencyii}
Directive 2009/138/{EC} of the {E}uropean {P}arliament and of the {C}ouncil of
  25 {N}ovember 2009 on the taking-up and pursuit of the business ({S}olvency
  {II}) of {I}nsurance and {R}einsurance.
\newblock Official Journal of the European Union, L351(1)  (2009)

\bibitem{asimit2019efficient}
Asimit, V., Peng, L., Wang, R., Yu, A.: An efficient approach to quantile
  capital allocation and sensitivity analysis.
\newblock Mathematical Finance \textbf{29}(4), 1131--1156 (2019)

\bibitem{baselMarketRisk}
BCBS: Minimum capital requirements for market risk.
\newblock Tech. rep., Bank for International Settlements (2016).
\newblock \urlprefix\url{http://www.bis.org/bcbs/publ/d352.pdf}

\bibitem{brownlees2012volatility}
Brownlees, C.T., Engle, R.: Volatility, correlation and tails for systemic risk
  measurement.
\newblock Available at SSRN \textbf{1611229} (2012)

\bibitem{buch2008coherent}
Buch, A., Dorfleitner, G.: {C}oherent risk measures, coherent capital
  allocations and the gradient allocation principle.
\newblock Insurance: Mathematics and Economics \textbf{42}(1), 235--242 (2008)

\bibitem{chacon2011asymptotics}
Chac{\'o}n, J.E., Duong, T., Wand, M.: Asymptotics for general multivariate
  kernel density derivative estimators.
\newblock Statistica Sinica pp. 807--840 (2011)

\bibitem{denault2001coherent}
Denault, M.: {C}oherent allocation of risk capital.
\newblock Journal of risk \textbf{4}(1), 1--34 (2001)

\bibitem{ksPackage}
Duong, T.: ks: Kernel Smoothing (2021).
\newblock \urlprefix\url{https://CRAN.R-project.org/package=ks}.
\newblock R package version 1.13.3

\bibitem{embrechts2014academic}
Embrechts, P., Puccetti, G., R{\"u}schendorf, L., Wang, R., Beleraj, A.: An
  academic response to basel 3.5.
\newblock Risks \textbf{2}(1), 25--48 (2014)

\bibitem{emmer2015best}
Emmer, S., Kratz, M., Tasche, D.: What is the best risk measure in practice?
  {A} comparison of standard measures.
\newblock Journal of Risk \textbf{18}(2), 31--60 (2015)

\bibitem{fan2020vector}
Fan, Y., Henry, M.: Vector copulas.
\newblock arXiv preprint arXiv:2009.06558  (2020)

\bibitem{fernandez1998bayesian}
Fern{\'a}ndez, C., Steel, M.F.: On bayesian modeling of fat tails and skewness.
\newblock Journal of the american statistical association \textbf{93}(441),
  359--371 (1998)

\bibitem{FournieMalliavin2}
Fourni\'e, E., Lasry, J.M., Lebuchoux, J., Lions, P.L.: Applications of
  {M}alliavin calculus to {M}onte-{C}arlo methods in finance. {II}.
\newblock Finance and Stochastics \textbf{5}(2) (2001)

\bibitem{fu2009conditional}
Fu, M.C., Hong, L.J., Hu, J.Q.: Conditional monte carlo estimation of quantile
  sensitivities.
\newblock Management Science \textbf{55}(12), 2019--2027 (2009)

\bibitem{glasserman2005measuring}
Glasserman, P.: {M}easuring marginal risk contributions in credit portfolios.
\newblock Journal of Computational Finance \textbf{9}(2), 1 (2005)

\bibitem{gourieroux2000sensitivity}
Gouri{\'e}roux, C., Laurent, J.P., Scaillet, O.: {S}ensitivity analysis of
  values at risk.
\newblock Journal of Empirical Finance \textbf{7}(3), 225--245 (2000)

\bibitem{hong2009estimating}
Hong, L.J.: Estimating quantile sensitivities.
\newblock Operations research \textbf{57}(1), 118--130 (2009)

\bibitem{iglehart1975simulating}
Iglehart, D.L.: Simulating stable stochastic systems, v: Comparison of ratio
  estimators.
\newblock Naval Research Logistics Quarterly \textbf{22}(3), 553--565 (1975)

\bibitem{kalkbrener2005axiomatic}
Kalkbrener, M.: {A}n axiomatic approach to capital allocation.
\newblock Mathematical Finance \textbf{15}(3), 425--437 (2005)

\bibitem{knothe1957contributions}
Knothe, H.: Contributions to the theory of convex bodies.
\newblock Michigan Mathematical Journal \textbf{4}(1), 39--52 (1957)

\bibitem{koike2020markov}
Koike, T., Hofert, M.: Markov chain monte carlo methods for estimating systemic
  risk allocations.
\newblock Risks \textbf{8}(1), 6 (2020)

\bibitem{koike2019estimation}
Koike, T., Minami, M.: Estimation of risk contributions with mcmc.
\newblock Quantitative Finance \textbf{19}(9), 1579--1597 (2019)

\bibitem{li2017gradient}
Li, Y., Turner, R.E.: Gradient estimators for implicit models.
\newblock In: International Conference on Learning Representations (2018)

\bibitem{liu2009kernel}
Liu, G., Hong, L.J.: Kernel estimation of quantile sensitivities.
\newblock Naval Research Logistics (NRL) \textbf{56}(6), 511--525 (2009)

\bibitem{mainik2014dependence}
Mainik, G., Schaanning, E.: On dependence consistency of covar and some other
  systemic risk measures.
\newblock Statistics \& Risk Modeling \textbf{31}(1), 49--77 (2014)

\bibitem{marshall1988families}
Marshall, A.W., Olkin, I.: Families of multivariate distributions.
\newblock Journal of the American statistical association \textbf{83}(403),
  834--841 (1988)

\bibitem{mcneil2015quantitative}
McNeil, A.J., Frey, R., Embrechts, P.: {Q}uantitative {R}isk {M}anagement:
  {C}oncepts, {T}echniques, and {T}ools.
\newblock Princeton {U}niversity {P}ress (2015)

\bibitem{nolan:2018}
Nolan, J.P.: Stable Distributions - Models for Heavy Tailed Data.
\newblock Birkhauser, Boston (2018).
\newblock In progress, Chapter 1 online at
  http://fs2.american.edu/jpnolan/www/stable/stable.html

\bibitem{peters2017bayesian}
Peters, G.W., Targino, R.S., W{\"u}thrich, M.V.: Bayesian modelling, monte
  carlo sampling and capital allocation of insurance risks.
\newblock Risks \textbf{5}(4), 53 (2017)

\bibitem{rosenblatt1952remarks}
Rosenblatt, M.: Remarks on a multivariate transformation.
\newblock The annals of mathematical statistics \textbf{23}(3), 470--472 (1952)

\bibitem{shi2018spectral}
Shi, J., Sun, S., Zhu, J.: A spectral approach to gradient estimation for
  implicit distributions.
\newblock In: International Conference on Machine Learning, pp. 4644--4653.
  PMLR (2018)

\bibitem{siller2013measuring}
Siller, T.: {M}easuring marginal risk contributions in credit portfolios.
\newblock Quantitative Finance \textbf{13}(12), 1915--1923 (2013)

\bibitem{silverman2018density}
Silverman, B.W.: Density estimation for statistics and data analysis.
\newblock Routledge (2018)

\bibitem{song2019generative}
Song, Y., Ermon, S.: Generative modeling by estimating gradients of the data
  distribution.
\newblock Advances in Neural Information Processing Systems \textbf{32} (2019)

\bibitem{sriperumbudur2017density}
Sriperumbudur, B., Fukumizu, K., Gretton, A., Hyv{\"a}rinen, A., Kumar, R.:
  Density estimation in infinite dimensional exponential families.
\newblock Journal of Machine Learning Research \textbf{18} (2017)

\bibitem{strathmann2015gradient}
Strathmann, H., Sejdinovic, D., Livingstone, S., Szabo, Z., Gretton, A.:
  Gradient-free hamiltonian monte carlo with efficient kernel exponential
  families.
\newblock Advances in Neural Information Processing Systems \textbf{28} (2015)

\bibitem{sutherland2018efficient}
Sutherland, D.J., Strathmann, H., Arbel, M., Gretton, A.: Efficient and
  principled score estimation with nystr{\"o}m kernel exponential families.
\newblock In: International Conference on Artificial Intelligence and
  Statistics, pp. 652--660. PMLR (2018)

\bibitem{targino2015sequential}
Targino, R.S., Peters, G.W., Shevchenko, P.V.: Sequential {M}onte {C}arlo
  samplers for capital allocation under copula-dependent risk models.
\newblock Insurance: Mathematics and Economics \textbf{61}, 206--226 (2015)

\bibitem{tasche1999risk}
Tasche, D.: {R}isk contributions and performance measurement.
\newblock Report of the Lehrstuhl f{\"u}r mathematische Statistik, TU
  M{\"u}nchen  (1999)

\bibitem{tasche2008capital}
Tasche, D.: {C}apital allocation to business units and sub-portfolios: the
  {E}uler principle.
\newblock In: Pillar II in the New Basel Accord: The Challenge of Economic
  Capital, pp. 423--453. Risk Books (2008)

\bibitem{zhou2020nonparametric}
Zhou, Y., Shi, J., Zhu, J.: Nonparametric score estimators.
\newblock In: International Conference on Machine Learning, pp. 11,513--11,522.
  PMLR (2020)

\end{thebibliography}

\begin{appendix}

\section{Proof of Theorem \ref{thm:extension}}\label{app:proof_main}

Following the approximation argument of the proof of \cite[Theorem 4.1]{FournieMalliavin2}, we consider the following formal identity:
$$\bE[X_i \ | \ g(\Y) = 0] = \frac{\bE[X_i \delta(g(\Y))]}{\bE[ \delta(g(\Y))]},$$
where $\delta$ is the Dirac delta at 0. Hence, if we define $\Delta(x) = \mathbbm{1}_{\{x \geq 0\}}$ and notice $\partial_j \Delta(g(\y))=\delta(g(\y)) \partial_j g(\y)$ for $j \in \{1,\dots,k\}$, then
$\delta(g(\y))=\partial_j \Delta(g(\y))/\partial_{j} g(\y)$
provided that
$\partial_jg(\y) \neq 0$.
This identity implies that
\begin{align*}
\bE[X_i\delta(g(\Y))]=\int_{\bR^k}g_i(\y)\delta(g(\y))f_\Y(\y)d\y=\int_{\bR^k}H_{ij}(\y) \partial_j\Delta(g(\y))d\y,
\end{align*}
where 
\begin{align*}
H_{ij}(\y) = \frac{g_i(\y)f_\Y(\y)}{\partial_j g(\y)}.
\end{align*}
Assume $j\neq i$ and let
\begin{align*}
B_{ij}(\y_{-j})=H_{ij}(\y)\Delta(g(\y))\Bigg|^{y_j=a_j^\text{Y,U}}_{y_j=a_j^\text{Y,L}}.
\end{align*}
Then integration by parts yields
\begin{align*}
\int_{\bR} H_{ij}(\y) \partial_j \Delta(g(\y))d y_j &= H_{ij}(\y)\Delta(g(\y))\Bigg|^{y_j=a_j^\text{Y,U}}_{y_j=a_j^\text{Y,L}}  - \int_{\bR} \Delta(g(\y))\partial_j H_{ij}(\y) d y_j \\
&= B_{ij}(\y_{-j})- \int_{\bR} \Delta(g(\y))\partial_j H_{ij}(\y) d y_j
\end{align*}
and thus
\begin{align*}
\bE[X_i\delta(g(\Y))]&=\int_{\bR^k}H_{ij}(\y)\partial_j\Delta(g(\y))d\y=\int_{\bR^{k-1}}
\left(\int_{\bR} H_{ij}(\y)\partial_j \Delta(g(\y))d y_j \right)
d \y_{-j}\\
&=\int_{\bR^{k-1}} B_{ij}(\y_{-j}) d \y_{-j}
 -\int_{\bR^k}  \Delta(g(\y))\partial_{j} H_{ij}(\y) d \y\\
 &=  B_{ij} +\int_{\bR^k} h_{ij}(\y) \Delta(g(\y)) d \y,
\end{align*}
where
\begin{align*}
 B_{ij}=\int_{\bR^{k-1}} B_{ij}(\y_{-j})d \y_{-j}\quad\text{and}
\quad
h_{ij}(\y)=-\partial_{j} H_{ij}(\y).
\end{align*}
Since
\begin{align*}
\frac{h_{ij}(\y)}{f_\Y(\y)}&= -\frac{\partial_j g_i(\y)}{\partial_j g(\y)} + g_i(\y) \left(\frac{\partial_j^2 g(\y)}{(\partial_j g(\y))^2} 
-\frac{\partial_{j}f_{\Y}(\y)}{f_{\Y}(\y)}\frac{1}{\partial_{j} g(\y)}
\right),
\end{align*}
we obtain
\begin{align*}
\bE[X_i\delta(\X)]&=  B_{ij} +\int_{\bR^k} h_{ij}(\y) \Delta(g(\y)) d \y,\\
&= B_{ij}+\bE[(X_i  \pi_j +  \pi_{ij})\Delta(g(\Y))]\\
&= B_{ij}+\bP(g(\Y) \geq 0)\bE[(X_i \pi_j + \pi_{ij}) \ | \ g(\Y) \geq 0],
\end{align*}
where
\begin{align*}
 \pi_j = \frac{\partial_j^2 g(\Y)}{(\partial_j g(\Y))^2} 
-\frac{\partial_{j}f_{\Y}(\Y)}{f_{\Y}(\Y)}\frac{1}{\partial_{j} g(\Y)}\quad
\mbox{ and }\quad  \pi_{ij} = - \frac{\partial_j g_i(\Y)}{\partial_j g(\Y)}.
\end{align*}
Similarly,
\begin{align*}
\bE[\delta(\X)]&=  B_j+\bP(g(\Y) \geq 0)\bE[ \pi_j \ | \ g(\Y) \geq 0],
\end{align*}
where
\begin{align*}
 B_{j}=\int_{\bR^{k-1}}\frac{f_\Y(\y_{-j},y_j)}{\partial_j g(\y_{-j},y_j)}\mathbbm{1}_{\{g(\y_{-j},y_j)\geq 0\}}\Bigg|^{y_j=a_j^\text{Y,U}}_{y_j=a_j^\text{Y,L}} d \y_{-j}.
\end{align*}
Consequently, we obtain the formula
\begin{align*}
\bE[X_i \ | \ g(\Y) = 0] &= \frac{ B_{ij} + \bP(g(\Y) \geq 0) \displaystyle \bE\left[X_i \  \pi_j +  \pi_{ij} \ | \ g(\Y) \geq 0 \right]}{ B_j + \bP(g(\Y) \geq 0))\displaystyle \bE\left[ \pi_j \ | \ g(\Y) \geq 0 \right]},
\end{align*}
for $j\neq i$.
Since the left-hand side is independent of $j$, we find
\begin{align*}
\bE[X_i \ | \ g(\Y) = 0] &= \frac{B_{iI} + \bP(g(\Y) \geq 0) \displaystyle \bE\left[X_i \ \pi_I + \pi_{iI} \ | \ g(\Y) \geq 0 \right]}{B_I + \bP(g(\Y) \geq 0)\displaystyle \bE\left[\pi_I \ | \ g(\Y) \geq 0 \right]},
\end{align*}
where $B_{iI} = \sum_{j\in I} B_{ij}$, $B_I=\sum_{j\in I} B_{j}$, $\pi_I = \sum_{j\in I} \pi_{j}$ and $\pi_{iI}=\sum_{j\in I} \pi_{ij}$. Finally, using the particular choice of $g$, we have the desired formula, since $\bP(g(\Y) \geq 0) = \bP(X \geq \VaR_\alpha(X)) = 1-\alpha$.

\end{appendix}

\end{document}